\documentclass[letterpaper,twocolumn,10pt]{article}
\usepackage{usenix-2020-09}

\usepackage{tikz}
\usepackage{amsmath}

\usepackage{algorithm}
\usepackage{array}
\usepackage{xfrac}
\usepackage{amssymb}
\usepackage{mathrsfs}
\usepackage{gensymb}
\usepackage{dsfont}
\usepackage{graphicx}
\usepackage{enumitem}
\usepackage{caption}
\usepackage[normalsize,TABBOTCAP]{subfigure}
\usepackage{comment}
\usepackage{xspace}
\usepackage{alphabeta}
\usepackage{array}
\usepackage{makecell}
\usepackage{xcolor}

\usepackage{url}

\newcommand{\ours}{\texttt{NR-Surface}\xspace}

\begin{document}

\date{}

\title{NR-Surface: NextG-ready $μ$W-reconfigurable mmWave Metasurface}

\author{
{\rm Minseok Kim$^{\dagger}$}\\
\and
{\rm Namjo Ahn$^{\dagger}$}\\
\\KAIST
\and
{\rm Song Min Kim$^{*}$}\\
} 

\maketitle

\begingroup\def\thefootnote{$\dagger$}\footnotetext{Co-primary student authors.

$\,~^{*}$Corresponding author. (songmin@kaist.ac.kr)}\endgroup
    
\begin{abstract}\label{abstract}

Metasurface has recently emerged as an economic solution to expand mmWave coverage. However, their pervasive deployment remains a challenge, mainly due to the difficulty in reaching the tight 260ns NR synchronization requirement and real-time wireless reconfiguration while maintaining multi-year battery life. This paper presents \ours, the first real-time reconfigurable metasurface fully compliant with the NR standard, operating at 242.7 $\mu$W for a 2.1-year lifetime on an AA battery. \ours incorporates (i) a new extremely low-power (14KHz sampling) reconfiguration interface, NarrowBand Packet Unit (NBPU), for synchronization and real-time reconfiguration, and (ii) a highly responsive and low-leakage metasurface designed for low-duty cycled operation, by carefully leveraging the structure and the periodicity of the NR beam management procedure in the NR standard. \ours is prototyped and evaluated end-to-end with NR BS built on srsRAN to demonstrate diverse usage scenarios including multiple \ours per BS, multiple UE per \ours, and 3D beamforming. Around-the-corner UE evaluations showcase \ours efficacy under different user mobility patterns (20.3dB gain) and dynamic blockage (22.2dB gain). 

\end{abstract}
    
\vspace{-3mm}
\section{Introduction}\label{sec:intro}
\vspace{-2mm}

With the rich bandwidth spanning multi-GHz, millimeter-wave (mmWave) is a key technology to breakthrough spectrum shortage in the sub-6GHz, towards enhanced mobile broadband (eMBB) under the exponential growth of mobile data in the 5G and 6G cellular services. However, the benefit of mmWave communication comes at the cost of the technical hurdle of substantial path loss -- limiting the coverage, especially under complex indoor scenarios with significant obstructions, and hindering reliability for mobile users. To address this, various approaches have been introduced, including dense deployment of base stations (BS)~\cite{NRsmallcellFumihide,NRsmallcell}, relays~\cite{ntontin2019relay, naqvi2021achieving, bi2022proximity}, and distributed antennas~\cite{mahboob2019fiber,sung20195g,moerman2022beyond,naqvi20215g}. While simple and straightforward, installing a large body of costly transceivers is a hurdle to pervasive deployment and widespread of mmWave technology.

More recently, metasurface has emerged as an economic solution to expand mmWave coverage. Diverse designs were introduced including reconfigurable metasurfaces that adapt to varying channel and user mobility~\cite{cho2021mmwall, cho2022towards}, and low-cost 3D printed metasurface~\cite{10.1145/3495243.3517024}, where they are commonly designed to operate by passively reflecting the BS signal without RF chain for signal generation. This significantly simplifies the metasurfaces' RF structure compared to the full-fledged mmWave transceiver for lower manufacturing costs. However, pervasive deployment of the metasurfaces still remains a challenge, mainly due to the lack of a metasurface synchronization mechanism (to NR infrastructure BS) and real-time wireless reconfiguration interface, especially with multi-year battery lifetime to minimize deployment and maintenance costs.

\begin{figure*}[t]
\centering
\includegraphics[width=2.1\columnwidth]{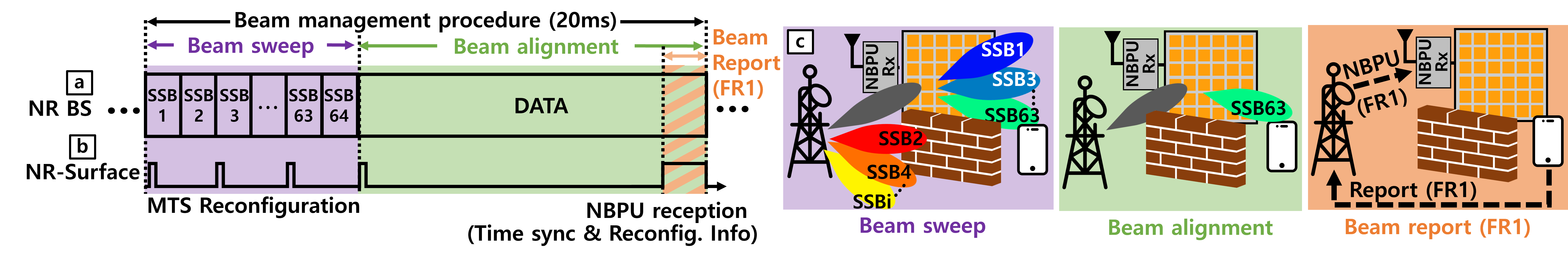}
\hspace{-11mm}
\vspace{-6mm}
\caption{(a) NR beam management procedure and (b) the corresponding \ours operation, offering NR-compliance. (c) Illustrative view of \ours operation in the beam management procedure.}
\label{fig:NR_background}
\hspace{-30mm}
\vspace{-7mm}
\end{figure*}

This paper presents \ours, the first real-time reconfigurable metasurface fully compliant with the NR standard operating at $\mu$W regime for a 2.1-year battery lifetime. For this \ours tackles the challenges imposed by the NR standard at only 242.7$\mu$W power: (i) 260ns timing synchronization with the BS~\cite{38104,38133} and (ii) in-time metasurface reconfiguration following the beam management procedure~\cite{38211,38212}. \ours co-designs extremely low-power (15KHz sampling rate) wireless reconfiguration interface and highly responsive and low-leakage varactor-based metasurface. The reconfiguration interface is overlayed on top of NB-IoT (on the NR guard band). This imposes no hardware change on the BS while minimizing disruption to the cellular traffic. \ours's metasurface achieves low-latency reconfiguration within the NR cyclic prefix (preventing communication error) with 10\% low-duty cycling. \ours performance is evaluated end-to-end with NR BS built on USRP X300 with srsRAN~\cite{srsRAN}, where it offers 20.3dB and 22.2dB gain for around-the-corner UEs under mobility and dynamic blockage, respectively. Support for multiple \ours per BS and multiple-UE per \ours were also evaluated for diversified practical scenarios.

\ours presents a reconfiguration interface of NarrowBand Packet Unit (NBPU), embedded on NR NB-IoT. NBPU is an asymmetric communication having NB-IoT with 180KHz bandwidth as the transmitter (BS side) and custom-built NBPU Rx with 15KHz bandwidth as the receiver (\ours side). BS carefully selects the NB-IoT symbol to be interpreted as an OOK NBPU symbol at the NBPU Rx, via aliasing and harmonics from the envelope detector. Equivalent time sampling is applied to achieve 260ns synchronization accuracy meeting the NR requirement, by leveraging the insight that the arrival timing of the NBPU symbol is easily inferable from the periodic nature of the NR beam management procedure (repeated every 20ms). NBPU also delivers timing and phase for metasurface reconfiguration. The beam management procedure offers another energy-saving opportunity, where the metasurface reconfiguration takes merely 3.8\% of the time. The metasurface simply holds the same configuration during the rest of the period. With varactor-based cells and GPIO-controlled control, \ours metasurface is designed for efficient low-duty cycling with 3ps low-latency reconfiguration and a negligible 280nW power leakage when the configuration is maintained.

As the first NR-controlled real-time reconfigurable mmWave metasurface, \ours co-designs a new low-duty-cycled metasurface and BS-controlled wireless interface to collaboratively achieve NR-compliance at $\mu$W power dissipation. We believe \ours is a practical step towards the pervasive deployment of mmWave technology in our everyday lives. Our contributions are as follows:

\vspace{-2mm}
\begin{itemize}

\item To the best of our knowledge, \ours is the first real-time reconfigurable, NR-compliant, and low-power mmWave metasurface towards pervasive deployment.

\vspace{-2mm}
\item \ours is immediately deployable and incurs a minimum cost, as it does not require any hardware change on the NR BS. 

\vspace{-2mm}
\item We prototype \ours on PCB (metasurface on Rogers RO4003C substrate) and perform diverse real-world evaluations under diversified usage settings, including multiple \ours, mobile users, and under dynamic blockages.

\end{itemize}
\vspace{-3mm}

\section{Background} \label{sec:background}
\vspace{-2mm}

\noindent \textbf{NR beam management procedure.} 
NR mmWave communication follows the beam management procedure depicted in Figure~\ref{fig:NR_background}(a) that incorporates the entire communication cycle -- beam sweep, report, and alignment (i.e., data communication). 
During beam sweep in the first 5ms of the period, the BS transmits up to 64 Synchronization Signal Blocks (SSB) with different beam patterns, one of which is selected by each UE and fed back to the BS via beam report.
Finally, the uplink and downlink data is exchanged during data beam alignment through the selected beam patterns. We note that the beam management procedure iterates exactly every 20ms and beam sweep, report, and alignment timings are known in advance. 
This offers the opportunity to integrate \ours into the procedure in low-power, saving energy by being active only at known duration in the beam management procedure.

While mmWave in NR is known for its high data rate, the link frequently suffers from low reliability due to beam misalignments and blockages, especially in dynamic and mobile scenarios. 
To alleviate this, NR features inter-band Carrier Aggregation (CA) that integrates the high-rate FR2 (mmWave) data plane with a robust FR1 control plane~\cite{GSMA_white}, as in existing 5G platforms including Qualcomm~\cite{QualComm_X65,QualComm_X70,QualComm_X75} and MediaTek modems~\cite{MediaTek_M80}.
This is used to integrate \ours control into the NR standard, where \ours is controlled via the robust FR1 (sub-6GHz) in an energy-efficient manner.

\vspace{1mm} \noindent \textbf{NB-IoT.} NB-IoT is an NR protocol for massive Machine Type Communication (mMTC) use cases, designed to serve IoT devices. To enable lightweight communication, NB-IoT uses the FR1 band and occupies only 180KHz bandwidth which is narrower than the minimum bandwidth of the underlying NR. This narrowband property allows NB-IoT to utilize the guardband of the FR1 band. Furthermore, NB-IoT only encodes data via 4QAM OFDM with 12 subcarriers, considering the limited performance of low-end devices. \ours leverages the NB-IoT guardband mode to integrate the control plane of \ours into the existing NR control plane.

\vspace{-2mm}

\begin{figure}[!h]
\centering
\vspace{-2mm}
\includegraphics[width=0.9\columnwidth]{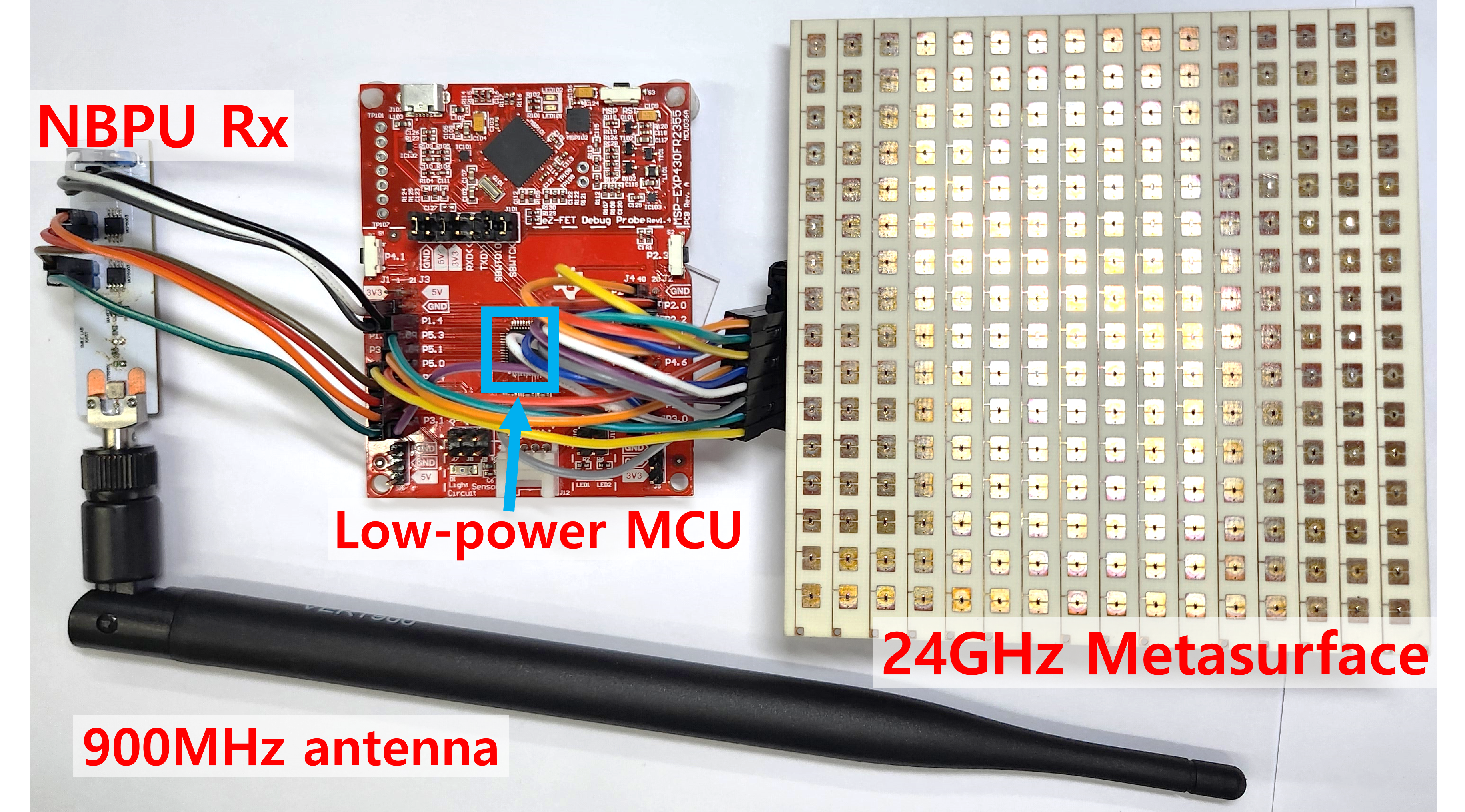}
\vspace{-3mm}	
\caption{\ours prototype implementation consisting of low-power reconfiguration interface with antenna, FR2 metasurface, and MCU. }\label{fig:overall_overview}
\vspace{-4mm}
\end{figure}

\vspace{-2mm}
\section{Design Overview}
\vspace{-3mm}

Figure~\ref{fig:overall_overview} illustrates \ours prototype, comprising 900MHz antenna, reconfiguration interface, low-power MCU, and 24GHz metasurface.
The 900MHz antenna captures the \ours's control channel, named NarrowBand Packet Unit (NBPU), embedded in the NR FR1. NBPU Rx synchronizes with the NBPU and decodes the reconfiguration info at $\mu$W-regime. The low-power MCU reconfigures the 24GHz metasurface via GPIO to steer the received beam based on the decoded info. 
Figure~\ref{fig:NR_background}(b) provides an overview of \ours operation aligned with the NR beam management procedure, explaining the illustrative view of \ours operation in Figure~\ref{fig:NR_background}(c).
During the beam sweep (Figure~\ref{fig:NR_background}(c) left), the 24GHz metasurface is reconfigured at the allocated SSBs, sweeping the beams beyond the blockages.
In the beam alignment (Figure~\ref{fig:NR_background}(c) middle), the metasurface steers the beam toward UE's direction as reported in the previous procedure.
NBPU Rx acquires reconfiguration info and timing for the next beam management procedure, based on the UE's beam report (Figure~\ref{fig:NR_background}(c) right).
During this whole procedure, our MCU enters idle mode to save energy when neither receiving NBPU nor reconfiguring the metasurface (Figure~\ref{fig:NR_background}(b)).

\ours design consists of two main parts enabling NR-compliant operation while at $\mu$W-consumption.
The first part, discussed in Section~\ref{sec:synchronization}, is the NBPU design for tight ns-synchronization with NR BS. To achieve this, the channel is co-designed with the $\mu$W-regime Rx to maximize the SNR and hence the synchronization accuracy, with signals selected to constructively add when passing through the low-end Rx. 
Section~\ref{sec:reconfiguration} presents a low-power metasurface design, tailored for real-time reconfigurations with short delays.
\ours implements duty-cycling to further save energy down to $242.7\mu$W, by hardware components designed to consume only $6.4\mu$W during the idle mode, maintaining configurations efficiently until known wake-up timings derived from the NR periodicity.

\vspace{-1mm}
\section{\texorpdfstring{$\mu$}{\textmu}W Synchronization with NR Base Station} \label{sec:synchronization}
\vspace{-1mm}

Per the NR standard~\cite{38104,38133}, a tight time synchronization of under $\pm 260$ns between BS and \ours is required to prevent FR2 communication errors. Doing so lets \ours reconfiguration fall within FR2 cyclic prefix (CP) of 585ns (i.e., 260$\times$2 with 65ns margin for multipath), thereby preventing error. For this \ours establishes an FR1 control channel with the BS for a tight and reliable synchronization and reconfiguration. To minimize the disruption on the ongoing NR traffic and maintain compatibility, this channel is built on NB-IoT on the FR1 guardband and readily available on NR BS. 

The control is a downlink-only (BS $\rightarrow$ \ours) channel carrying the synchronization (i.e., preamble) and reconfiguration info, where BS uses NB-IoT for Tx. Meanwhile, \ours aims at $\mu$W-regime and is unable to equip NB-IoT Rx that consumes 75mW~\cite{nbiot_power} to limit the battery life to only two days on a single AA~\cite{aa_battery}. For this we present NarrowBand Packet Unit (NBPU), an asymmetric control channel overlayed on NB-IoT. In particular, by emulating with NB-IoT, NBPU is designed as a single-carrier 15KHz OOK. \ours uniquely leverages aliasing from the bandwidth asymmetry between the 180KHz NBPU Tx (i.e., NB-IoT in the BS) and 15KHz NBPU Rx, to achieve 234ns synchronization (therefore meeting that NR requirement) and decoding with only 119.3$\mu$W for 2.1-year battery lifetime with a single AA.

\begin{figure}[h!]
\centering
\includegraphics[width=\columnwidth]{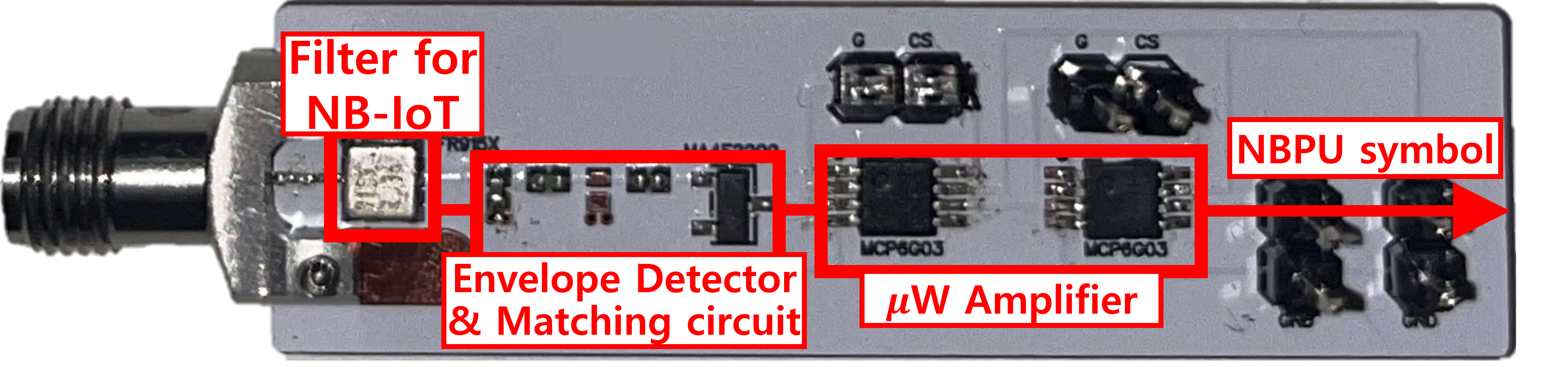}
\vspace{-2mm}
\caption{NBPU Rx to yield NBPU symbol from NB-IoT, comprising a filter for NR guardband operation, envelop detector for zero-power downconversion, and $\mu$W amplifiers to boost SNR.} \label{fig:Rx_hardware}
\vspace{-5mm}
\end{figure}

\begin{figure}[h!]
\centering
\vspace{-2mm}
\subfigure{\raisebox{1.75mm}{\includegraphics[width=0.5\columnwidth]{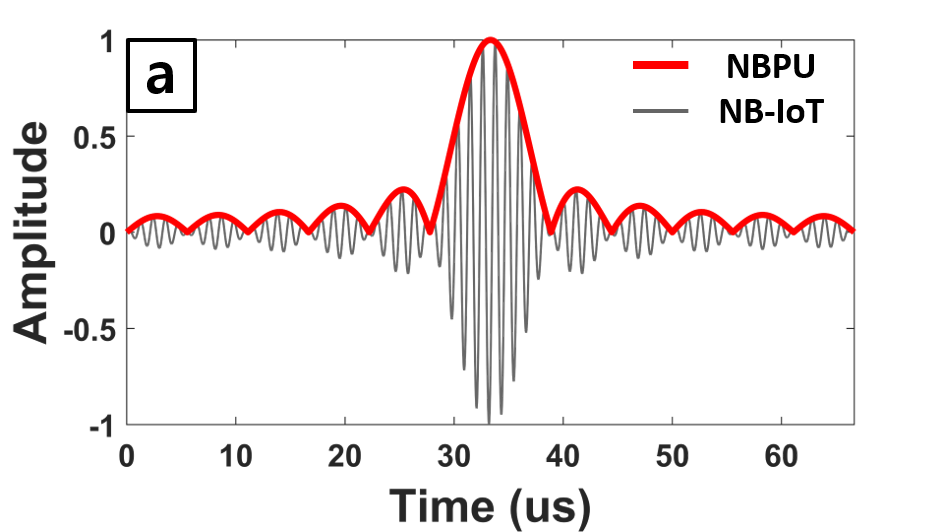}}}
\subfigure{\includegraphics[width=0.455\columnwidth]{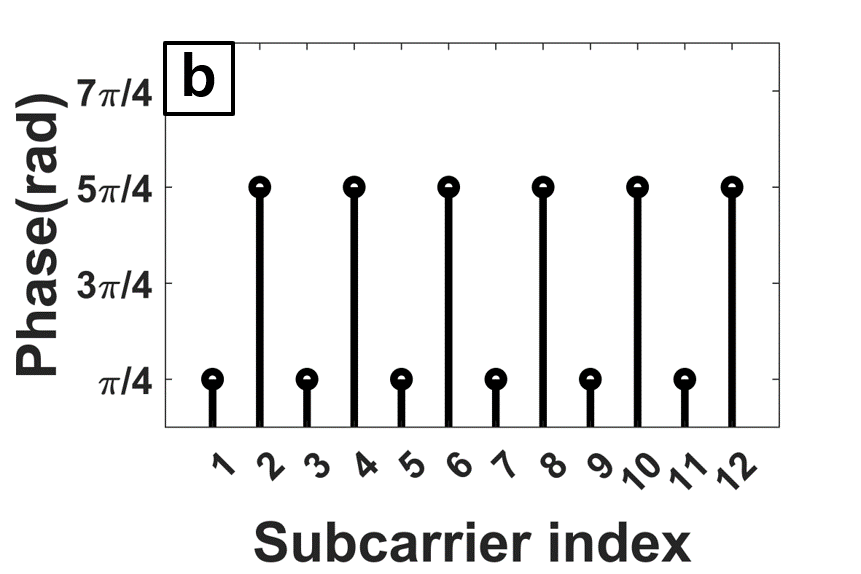}}
\hspace{1mm}
\vspace{-4mm}	
\caption{(a) The selected NBPU symbol with the maximum SNR and the corresponding (b) NB-IoT symbol (4QAM OFDM with 12 subcarriers).}
\label{fig:SNR_max_signal}
\vspace{-0.3cm}
\end{figure}

\vspace{-0.2cm}
\subsection{NBPU Embedding in NB-IoT}

\noindent\textbf{NBPU Symbol.}
Here we discuss generating an NBPU symbol from an NB-IoT symbol -- 12 subcarrier 4QAM OFDM with 15KHz spacing (from 15KHz to 180KHz). NBPU Rx in Figure~\ref{fig:Rx_hardware} leverages the envelope detector for zero-power downconversion. The envelope comprises multiple harmonics, where the 1st order harmonics dominate over higher order harmonics ($\sim$37.3dB in our experiments). From this, the envelope is modeled entirely as the set of 1st-order harmonics -- Each component has the frequency and phase from the difference between a pair of NB-IoT subcarriers, where the envelope detector yields the summation of all harmonic components from all subcarriers pairs. Letting $\Delta f$ denote 15KHz subcarrier spacing and $\phi$ represent one of 4QAM phases, $\left\{\frac{\pi}{4}, \frac{3\pi}{4}, \frac{5\pi}{4}, \frac{7\pi}{4}\right\}$, the envelope becomes $\sum_{k=1}^{11} \sum_{i=k+1}^{12} \mathrm{cos}(2\pi k\Delta f t ~+~ \phi_i - \phi_{i-k})$ where $k\Delta f$ and $(\phi_{i}-\phi_{i-k})$ denote the frequency and phase difference between $i$th and $i-k$th subcarriers. This indicates that the phase of the envelope is determined solely by the combinations of the NB-IoT subcarrier phases.

Figure~\ref{fig:SNR_max_signal} depicts the NBPU symbol and the corresponding NB-IoT symbol. Among all possible NB-IoT envelopes, we use the one with the maximum power as the NBPU symbol for the maximum SNR and thus, the best synchronization performance. This is achieved by selecting an NB-IoT symbol such that its 1st harmonics (i.e., subcarrier difference from the envelope detector) with equal frequencies align in phase; All 1st harmonics are constructively added in the frequency domain (and in the time domain from the Parseval's theorem) to yield the maximum power. 
To see why, without loss of generality, let all odd number subcarriers have phases $\pi/4$ and the other subcarriers have phases $5\pi/4$ (as shown in Figure~\ref{fig:SNR_max_signal}(b)). This makes harmonics' phases $k\pi$ for $2\pi k\Delta f$ frequency component.
These harmonics yield an envelope of $\sum_{k=1}^{11} (12-k)\mathrm{cos}(2\pi k\Delta f t+ k\pi$). Here, the amplitude of each frequency component, $k\Delta f$, is maximized to $(12-k)$, the sum of all harmonics. Figure~\ref{fig:SNR_max_signal}(a) illustrates the NBPU symbol and its NB-IoT symbol whose 12 subcarrier phases are carefully chosen.\footnote{Implementing the target NB-IoT subcarrier phases requires careful payload selection and reverse engineering of the channel coding. Details are in Appendix~\hyperref[sec:appendix_a]{ A}.} As shown in Figure~\ref{fig:Rx_hardware}, the NBPU symbol is extracted with the implemented NBPU Rx board, which consists of the filter for extracting NB-IoT band, the envelope detector for downconversion, and $34\mu$W consuming amplifier.

\vspace{-1mm}
\begin{figure}[h!]
\centering
\includegraphics[width=\columnwidth]{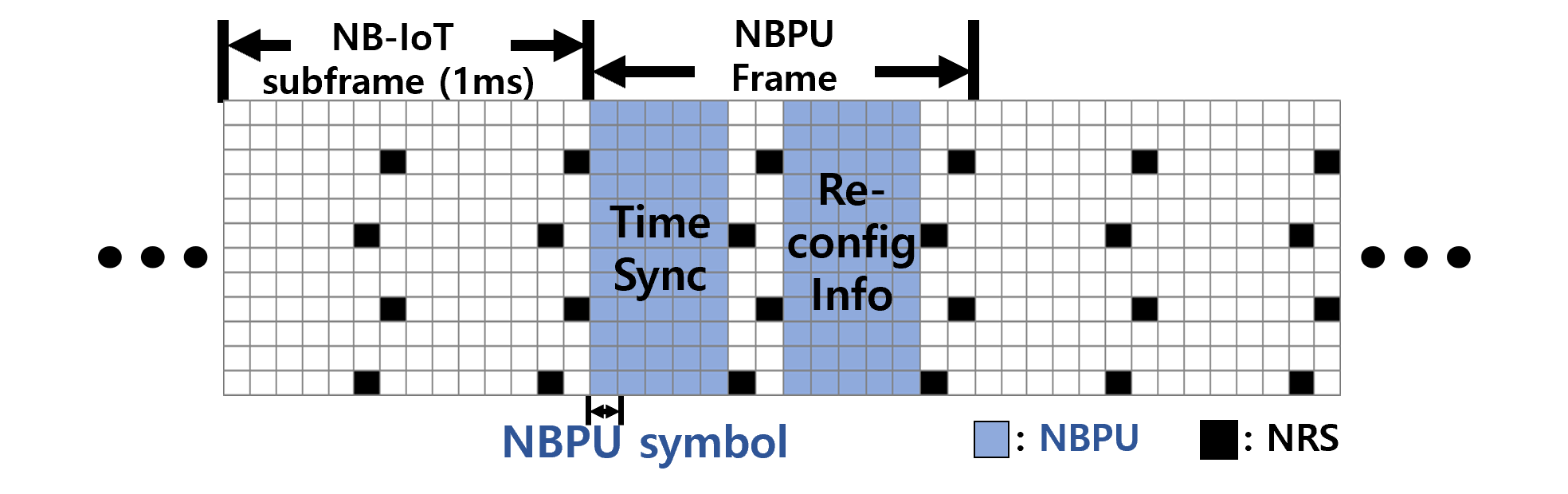}
\vspace{-6mm}
\caption{An NBPU Frame fits into one NB-IoT subframe. It has sync and reconfiguration info. (i.e., timing and phase).}
\label{fig:NBPU_frame}
\vspace{-0.2cm}
\end{figure} 

\noindent\textbf{NBPU Frame.} As depicted in Figure~\ref{fig:NBPU_frame} an NBPU frame is fit into a single NB-IoT subframe. An NBPU frame involves 5 NBPU symbols for synchronization and another 5 for the reconfiguration info, where two columns following the synchronization and reconfiguration info are unused as they are preoccupied with the mandatory Narrowband Reference Signal (NRS) for channel sounding. 
We note that one NBPU frame is inserted every 20 NB-IoT subframes to align with the periodic NR beam management procedure repeating every 20ms (Details in Section~\ref{sec:reconfiguration}).

\begin{figure}[h!]
\centering
\hspace{-5mm}
\includegraphics[width=0.9\columnwidth]{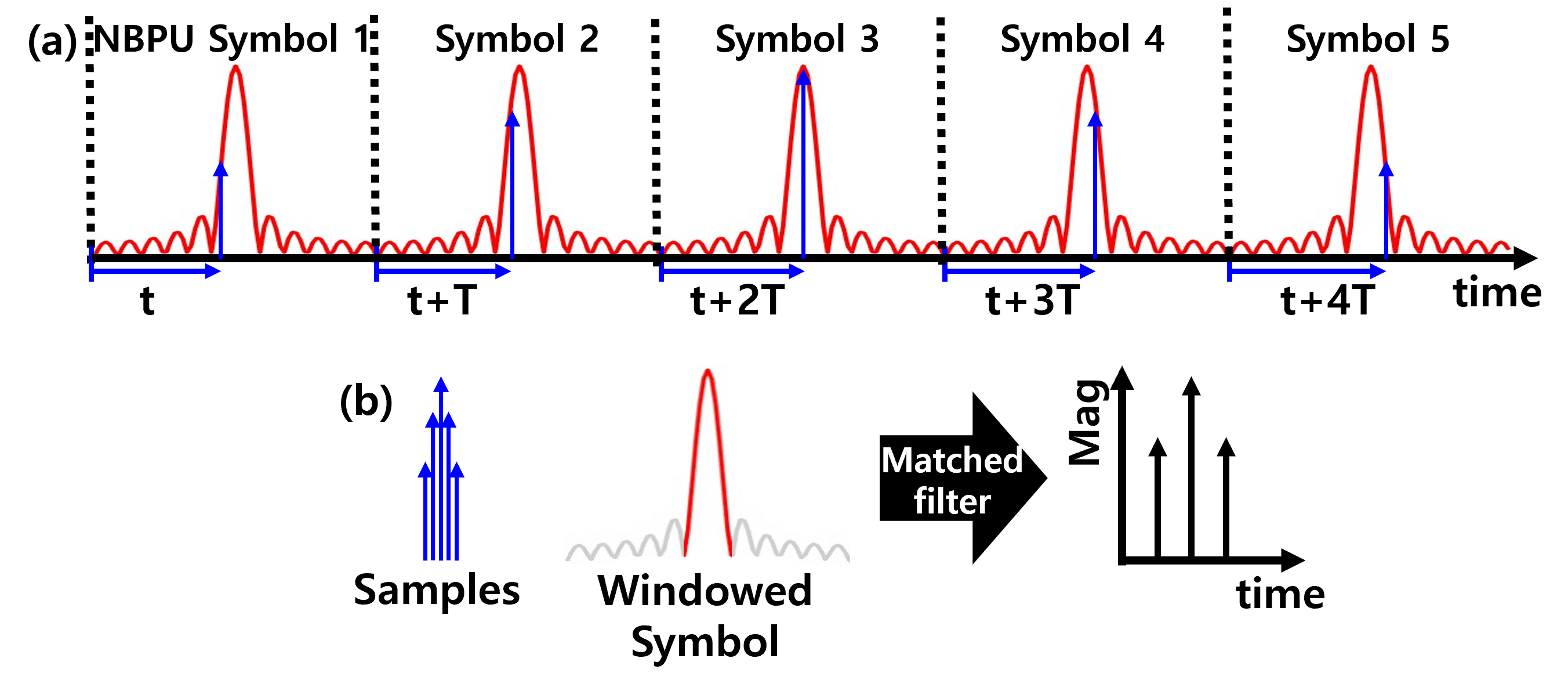}
\hspace{-7mm}
\vspace{-4mm}
\caption{(a) Equivalent time sampling utilizes the accumulated sampling time offset $T$ where $t$ is common offset for all five symbols. (b) The windowed symbol makes the computationally complex matched-filtering light.}
\label{fig:STO_effect}
\vspace{-5mm}
\end{figure}

\subsection{\texorpdfstring{$ns$}{ns}-synchronization}
Meeting the NR requirement of $\pm 260$ns synchronization accuracy requires the corresponding sampling rate of 3.84Msps~\cite{proakis2013}. However, running the ADC at this rate consumes mW-level power in low-power SAR ADCs~\cite{ADI_ADC, TI_ADC}, which is unaffordable for \ours. Instead, \ours leverages repeated symbols from which offset ($T$ in Figure~\ref{fig:STO_effect}) samples mimic the higher sampling rate, or equivalently, obtain fine-grained samples -- a technique known as the Equivalent-Time Sampling~\cite{time_equivalent_sampling}. Specifically, as in Figure~\ref{fig:STO_effect}(a), five NBPU symbols are repeated back-to-back with a single sample per symbol (14Ksps sampling rate); By enforcing an offset of $T=260$ns between consecutive samples (using MCU clock), the five samples collectively mimic a 3.84Msps sampling with only 14Ksps ADC -- cutting down the ADC power consumption by $274\times$ compared to using 3.84Msps ADC.

\ours runs matched filtering on the samples for optimal detection and synchronization~\cite{proakis2013}.
In particular, as in Figure~\ref{fig:STO_effect}(b), matched filtering is performed only on a central 780ns window out of 66.7$\mu$s symbol. This (i) significantly reduces the computation and energy consumption, and (ii) maintains the synchronization performance as the signal power is concentrated in the central portion of the symbol.
This requires first finding the center of the symbol so as to sample within the window -- Upon system bootup, \ours simply runs Equivalent-Time Sampling across the entire symbol from which the center is found by matched filtering. This takes 1.2s to complete (derivation in Appendix~\hyperref[sec:appendix_b]{ B}) and is a one-time task as the NBPU Time Sync repeats exactly every 20ms following the beam management protocol in NR standard.
\ours achieves 234ns average synchronization error under 0dB SNR (>30m~\cite{38901}), as evaluated in Section~\ref{subsection:sync_performance}.

\begin{figure}[h!]
\centering
\includegraphics[width=\columnwidth]{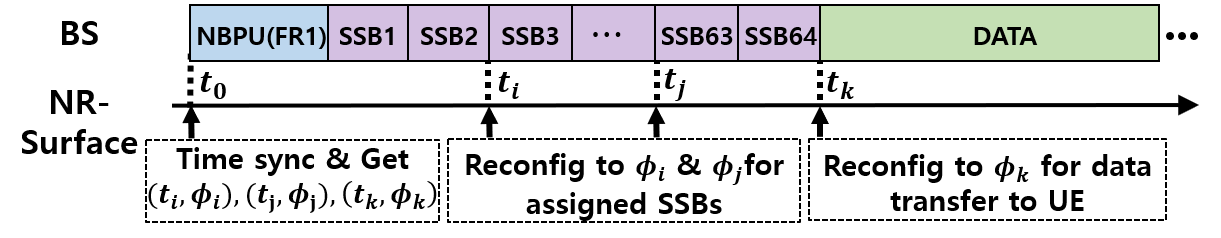}
\hspace{1mm}
\vspace{-7mm}
\caption{Interoperation overview to control FR2. BS transmits NBPU to synchronize $t_0$ and inform reconfiguration such as $(t_i,\phi_i),(t_j,\phi_j),(t_k,\phi_k)$. Then, \ours reconfigures in the right time and beam.}
\label{fig:NBPU_overview}
\hspace{-30mm}
\vspace{-7mm}
\end{figure}

\begin{figure}[h!]
\centering
\includegraphics[width=0.9\columnwidth]{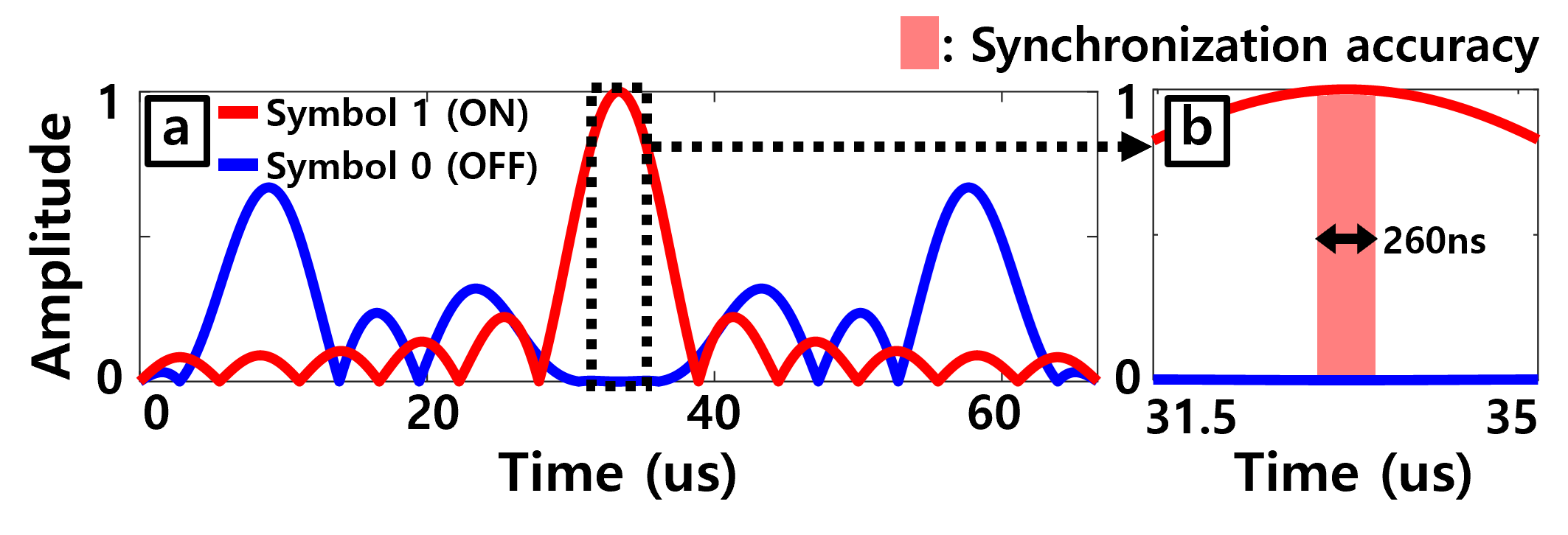}
\hspace{1mm}
\vspace{-5mm}
\caption{(a) Two NBPU symbols indicating 0 and 1. (b) They are OOK symbol pairs within the synchronization accuracy (red box).}
\label{fig:NBPU_symbols}
\hspace{-30mm}
\vspace{-7mm}
\end{figure}

\vspace{-3mm}
\section{\texorpdfstring{$\mu$}{\textmu}W Real-time Reconfiguration} \label{sec:reconfiguration}
\vspace{-1mm}

\ours needs to be tightly coupled with the NR beam management procedure for interoperability.
As shown in Figure~\ref{fig:NBPU_overview}, upon receiving the NBPU frame, \ours performs synchronization to obtain $t_0$ and extracts reconfiguration info, $(t_i,\phi_i),(t_j,\phi_j),(t_k,\phi_k)$. Including one NBPU symbol used for synchronization, We design two NBPU symbols (illustrated in Figure~\ref{fig:NBPU_symbols}(a)), enabling low-power and robust communication of reconfiguration info.
Specifically, two NBPU symbols are interpreted as 15KHz OOK symbol pairs by NBPU Rx, and our synchronization accuracy allows us to sample only the central part (red box in Figure~\ref{fig:NBPU_symbols}(b)).\footnote{Since NB-IoT can be turned on/off in the subframe scale, the NBPU symbol is needed to represent OFF state, whose NB-IoT subcarrier phase is $[\pi/4, \pi/4, 5\pi/4, 5\pi/4, 5\pi/4, 5\pi/4, \pi/4, \pi/4, \pi/4, \pi/4, 5\pi/4, 5\pi/4]$.}
This asymmetric communication (180KHz $\rightarrow$ 15KHz) allows NBPU Rx to focus on only the constructively/destructively added portion of each NBPU symbol.
This offers an additional 13dB SNR gain, which helps robust communication even with the passive envelope detector.

\begin{figure}[h!]
\centering
\includegraphics[width=\columnwidth]{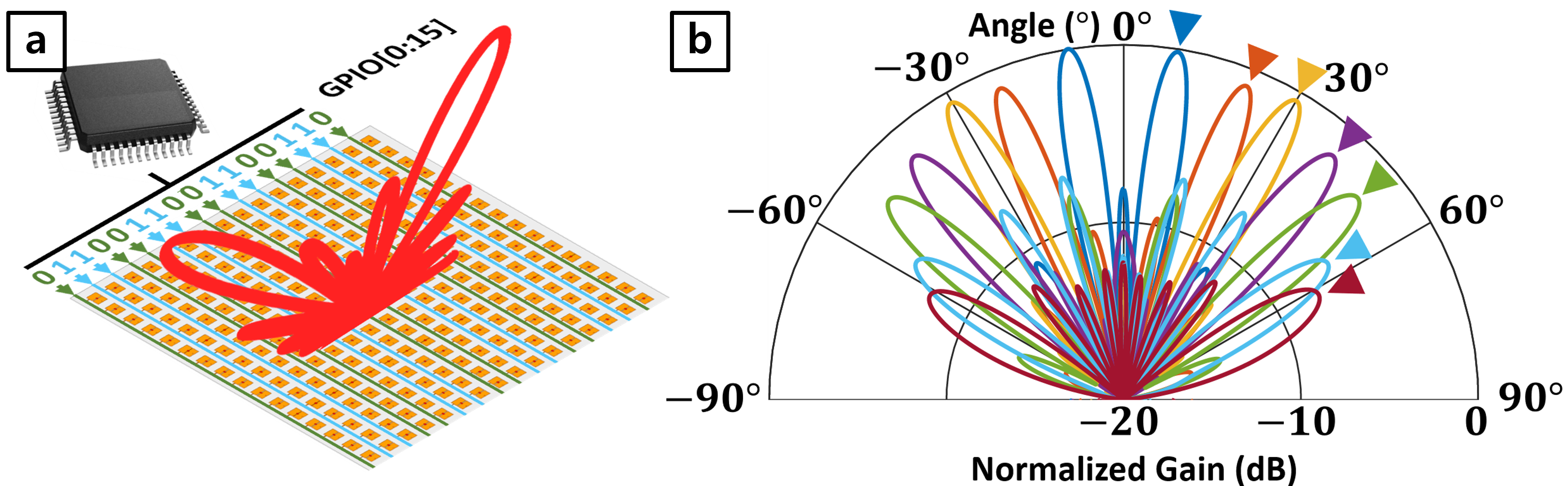}
\vspace{-0.6cm}
\caption{(a) GPIO-controlled metasurface structure and (b) example beam patterns. The main lobes are marked with triangles.} \label{fig:metasurface_beam}
\vspace{-0.6cm}
\end{figure}

\subsection{\texorpdfstring{$ns$}{ns}-reconfigurable Metasurface}

Figure~\ref{fig:metasurface_beam}(a) illustrates the metasurface consisting of a 16x16 array of unit-cells, reconfigured in real-time at NR-synchronized timings derived from NBPU. 
In satisfying the NR synchronization requirement, the reconfiguration should finish within the margin between the NR requirement and \ours accuracy.
\ours achieves this with the varactor-based unit-cell design and 1-bit digital output (GPIO) control, whose short response time allows \ours to reconfigure beam patterns in less than $10$ns while consuming only negligible energy ($\sim12$pJ in our implementation). 
From this, \ours forms narrow beam patterns towards various directions to cover a wide area. Figure~\ref{fig:metasurface_beam}(b) plots example beam patterns that \ours provides with 16 columns, whose half-power beam width is $6.3\degree$.
Moreover, $<260$ns NR synchronization requirement is still satisfied at a low SNR of $+0$dB, including both the synchronization error from NBPU and reconfiguration delay.

\begin{figure}[h!]
\centering
\vspace{-2mm}
\includegraphics[width=\columnwidth]{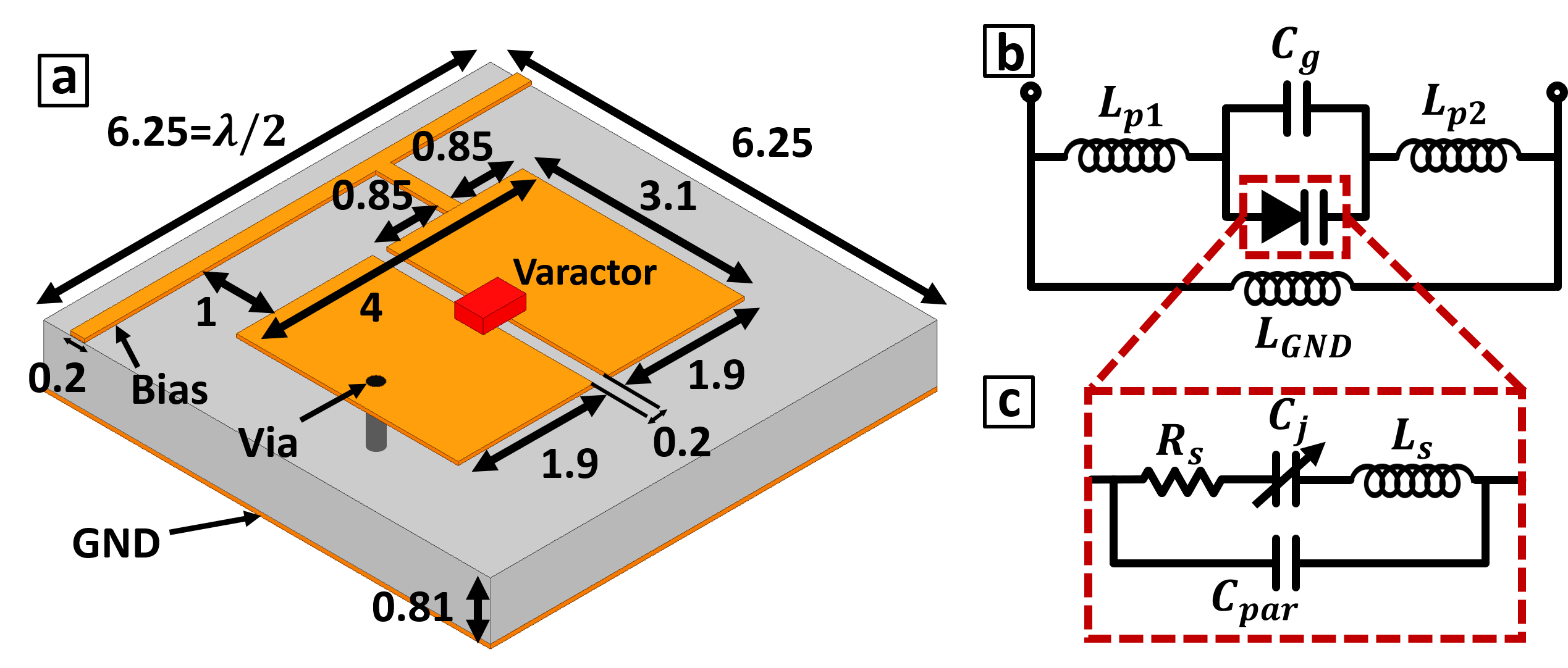}
\vspace{-6mm}
\caption{\ours (a) unit-cell design and equivalent circuit models for (b) unit-cell and (c) varactor diode. Dimensions are in mm.} \label{fig:unit_cell_design}
\vspace{-0.1cm}
\end{figure}

Figure~\ref{fig:unit_cell_design}(a) illustrates \ours unit-cell design achieving 1-bit phase shift in 10ns while consuming only $\sim12$pJ.
Unit-cell consists of two metal plates and a single varactor diode (Figure~\ref{fig:unit_cell_design}(b)), whose rectangular structure has been shown to be effective as a phase shifter~\cite{tian2017design, song2022reconfigurable}. 
Specifically, the MAVR-011020-1411 varactor diode~\cite{varactor_diode} is placed, consisting of a varying capacitance from $C_j=0.22$pF to $C_j=0.04$pF at $0$V to $15$V bias, respectively, with a series resistance of $R_s=13\Omega$ (Figure~\ref{fig:unit_cell_design}(c)). 
The GPIO interface is directly connected to provide bias voltage across the diode, allowing fast reconfigurations within 10ns owing to short rise/fall times of the GPIO interface ($<7$ns in our implementation) and the small time constant $R_sC_j=3$ps of the diode. Furthermore, the unit-cell only consumes $\sim12$pJ during reconfiguration, mainly due to small energy stored and dissipated at the varactor diode.
However, the reduced voltage range of GPIO limits the minimum capacitance of the diode to $C_j=0.08$pF at $3.3V$. Hence, the key design parameters such as length, width, and gap of metal plates are chosen, such that the varactor diode in parallel still shifts the phase by $180\degree$ over the reduced capacitance range.

\begin{figure}[h!]
\centering
\vspace{-4mm}
\hspace{-1mm}
\subfigure {\includegraphics[width=0.46\columnwidth]{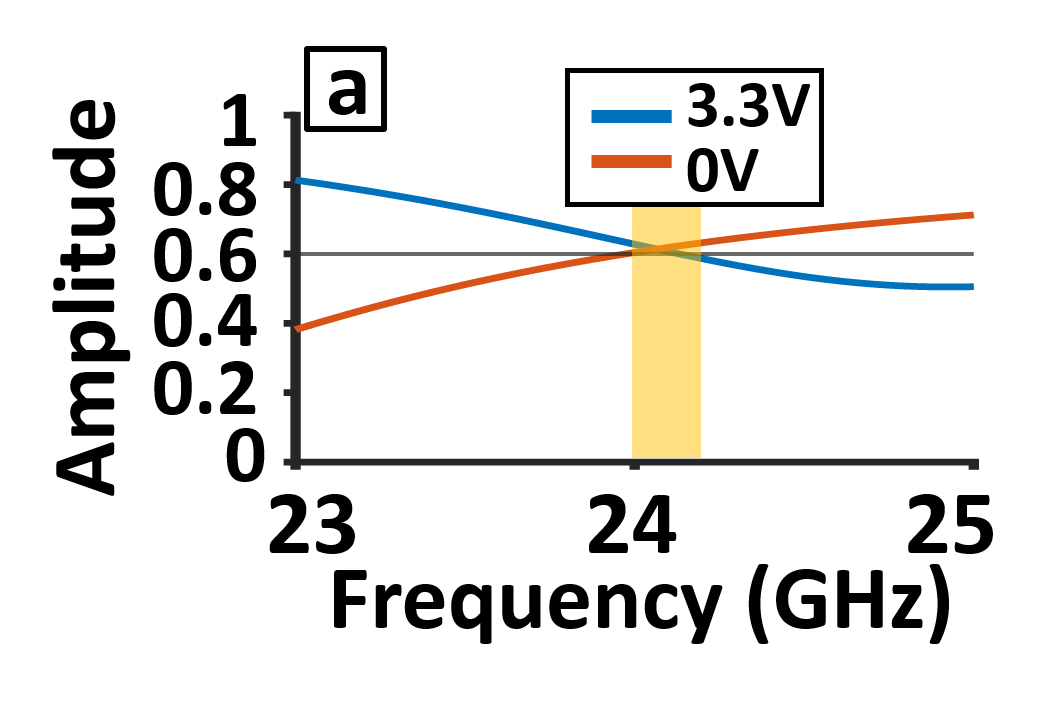}}
\hspace{2mm}
\subfigure {\includegraphics[width=0.46\columnwidth]{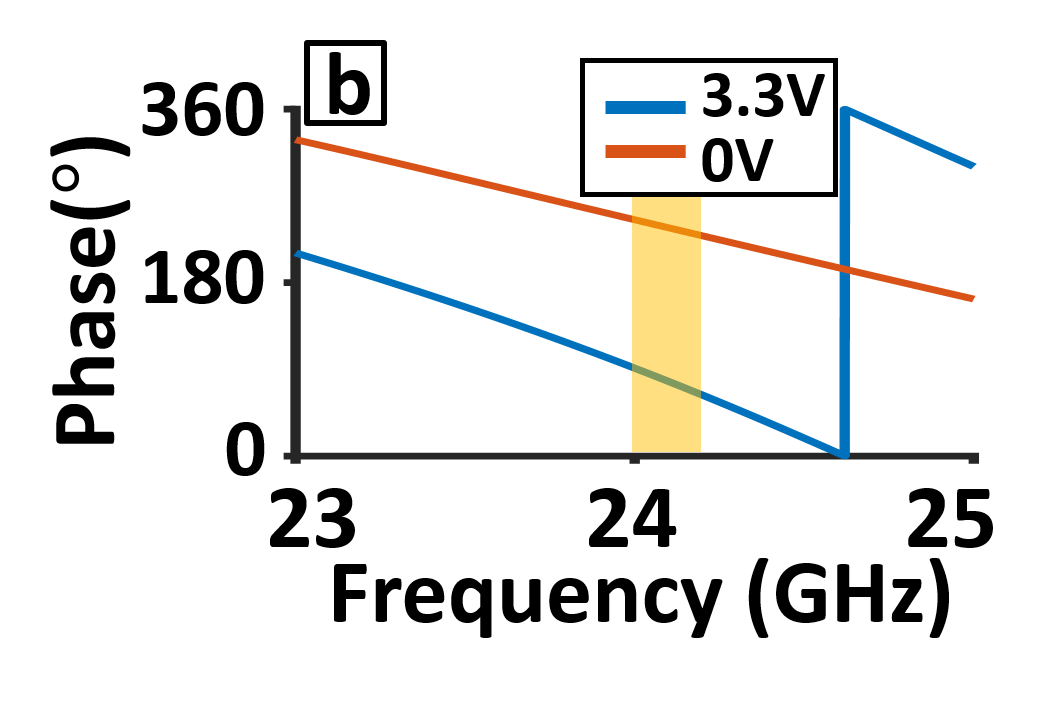}}
\vspace{-6mm}	
\caption{(a) Amplitude and (b) phase of the reflection coefficient when 0V and 3.3V are applied.} \label{fig:reflection_coefficient}
\vspace{-0.1cm}
\end{figure}

\begin{figure*}[t!]
\centering
\vspace{-2mm}
\includegraphics[width=2.1\columnwidth]{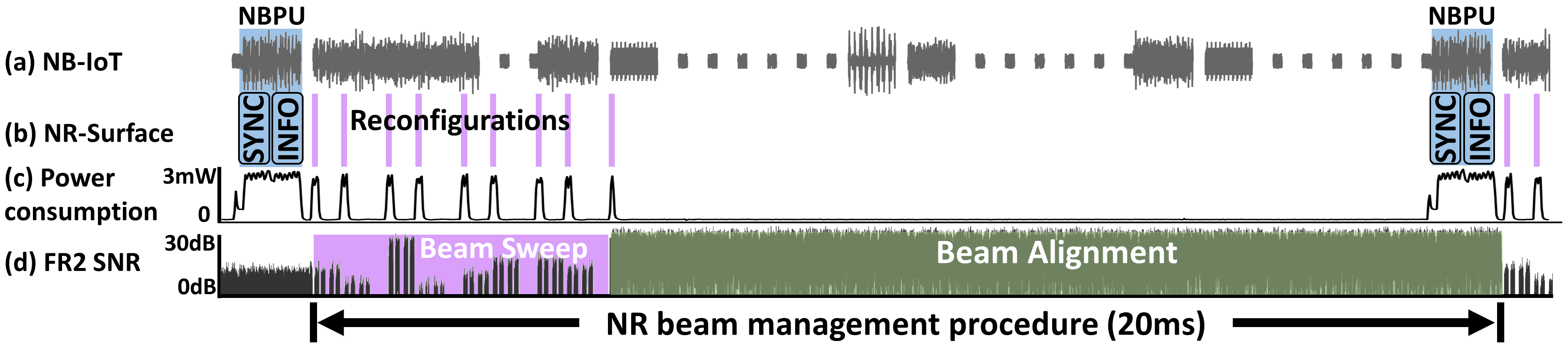}
\hspace{1mm}
\vspace{-6mm}	
\caption{Overall operation of \ours. (a) NB-IoT containing the NBPU frame, (b) \ours duty-cycled operation synchronized with NR BS, (c) Measured power consumption, and (d) FR2 signal with improved SNR from aligned beams.} \label{fig:duty_cycle_reconfiguration}
\vspace{-4mm}
\end{figure*}

Figure~\ref{fig:unit_cell_design}(a) shows the fine-tuned dimensions aimed at 1-bit phase shift, from simulations conducted with Ansys HFSS~\cite{HFSS} on the RO4003C substrate with $\epsilon_r=3.55$. The amplitude and phase of the reflection coefficient are plotted in Figure~\ref{fig:reflection_coefficient}(a) and (b), respectively, showing \ours unit-cell design achieves $\sim180\degree$ phase shift with $0.6$ amplitude across the entire 200MHz channel.
We note the simulation and the implementation are carried out on 24GHz ISM band which shares similar RF properties with 28GHz NR mmWave, for the prototype evaluation.
Figure~\ref{fig:metasurface_beam}(b) plots the example beam patterns simulated with HFSS, whose main lobe is steered towards different directions from $10\degree$ to $70\degree$ from the normal direction.
The beam width in the example is $6.3\degree$, however, this can be as narrow as $1.3\degree$ for 80-column metasurface, leveraging large availability of GPIO (up to 80 interfaces in typical low-power MCUs~\cite{MSP430, STM8, Microchip8}).
We note \ours can also be extended to provide 3D beam patterns as evaluated in Section~\ref{sec:3D_beampattern} with separately controlled unit-cells.
To this end, \ours reconfigures beam patterns towards a wide range of $\pm70\degree$, where 256 unit-cells consume only a negligible amount of $\sim3nJ$. 
While these energy-efficient reconfigurations facilitate low-power operation, the rest of the period should also be made low-power as well since ns-reconfigurations occupy a small portion of the period.

\vspace{-4mm}
\subsection{Duty-cycling \ours}
\vspace{-1mm}

The final piece to $\mu$W operation is NR-synchronized low duty-cycling, by leveraging the two opportunities from the beam management procedure: (i) The synchronization and reconfiguration only take 10\% and \ours is kept idle with the beam pattern maintained for the rest of the time. Also, (ii) the idle/wakeup times are known a priori, as beam management procedure repeats every 20ms. This makes in-time mode switch possible even in \ours's low-end hardware with slow wake up (e.g., $10\mu$s for MCU). In particular, all hardware components (i.e., NBPU Rx, MCU, and metasurface) are carefully designed to collaboratively achieve idle power consumption of 6.4$\mu$W. Specifically, the varactor-based unit-cell maintains the phase at an extremely low 72nW leakage. In the meantime, the bias voltage via GPIO is kept stable under low power modes supported by typical MCUs at hundreds of nW range~\cite{MSP430, STM8, Microchip8} and NBPU Rx is turned off. The MCU and NBPU Rx wakes up only at the time when the NBPU frame is expected.

\vspace{-3mm}
\subsection{Putting It All Together} 
\vspace{-1mm}

With duty-cycling, $\mu$W synchronization, and reconfiguration \ours achieves $242.7\mu$W consumption for a 2.1-year lifetime on a single AA battery.
Figure~\ref{fig:duty_cycle_reconfiguration} plots the overall operation from the experiment results encompassing synchronization, reconfiguration, and beam alignment, as well as the corresponding power consumption. In Figure~\ref{fig:duty_cycle_reconfiguration}(a), (b) \ours synchronizes with NBPU and decodes reconfiguration info. As shown in Figure~\ref{fig:duty_cycle_reconfiguration}(c) this step involves the heaviest computation to consume near half of the entire power ($119.3\mu$W). Metasurface reconfiguration and the corresponding short power bursts in Figure~\ref{fig:duty_cycle_reconfiguration}(b) and (c), respectively, take place in accordance with beam sweep and alignment in Figure~\ref{fig:duty_cycle_reconfiguration}(d). The short bursts sum up to 117$\mu$W. 
\ours then keeps the beam towards the UE for the rest of the period. This is when the duty-cycling is in effect, consuming only $6.4\mu$W to maintain the metasurface configuration. 
We note \ours is reconfigured only at half-subframes for non-colocated scenarios (i.e., NB-IoT and FR2 BS in separate devices) which increments synchronization error by $260$ns, leveraging longer CPs of 1106ns at half-subframe boundaries to avoid signal losses.
Collectively, battery-powered operation facilitates pervasive deployment without wall plugs, in dynamic indoor scenarios as well as outdoors.
\ours real-time operations in dynamic environments are evaluated in Section~\ref{subsec:dynamic_environment_evaluation}.

\vspace{-3mm}
\subsection{\textbf{Multi \ours}}
\vspace{-1mm}

Multiple \ours for a single BS may be necessary under complex environments with a large number of blockages and blind spots. For this, a separate set of SSBs is allocated to each \ours and the same NR-compliant operation is applied.
The BS tracks which SSB (i.e., beam pattern) corresponds to which \ours, and transmits multiple NBPUs for independent reconfiguration. 
In particular, they are embedded in different frames of a single NB-IoT spectrum, or extended to multiple NB-IoT spectrums in other NR guardbands for a large number of deployments. Beam patterns of up to 64 limit the number of \ours that can cooperate with a single BS, however, exceeding beam patterns can be leveraged by other BS to benefit from dense \ours deployment. We evaluated the real-time multiple operations in Section~\ref{subsec:multi_eval}.

\begin{figure}[h!]
\centering
\vspace{-2mm}{\includegraphics[width=0.9\columnwidth]{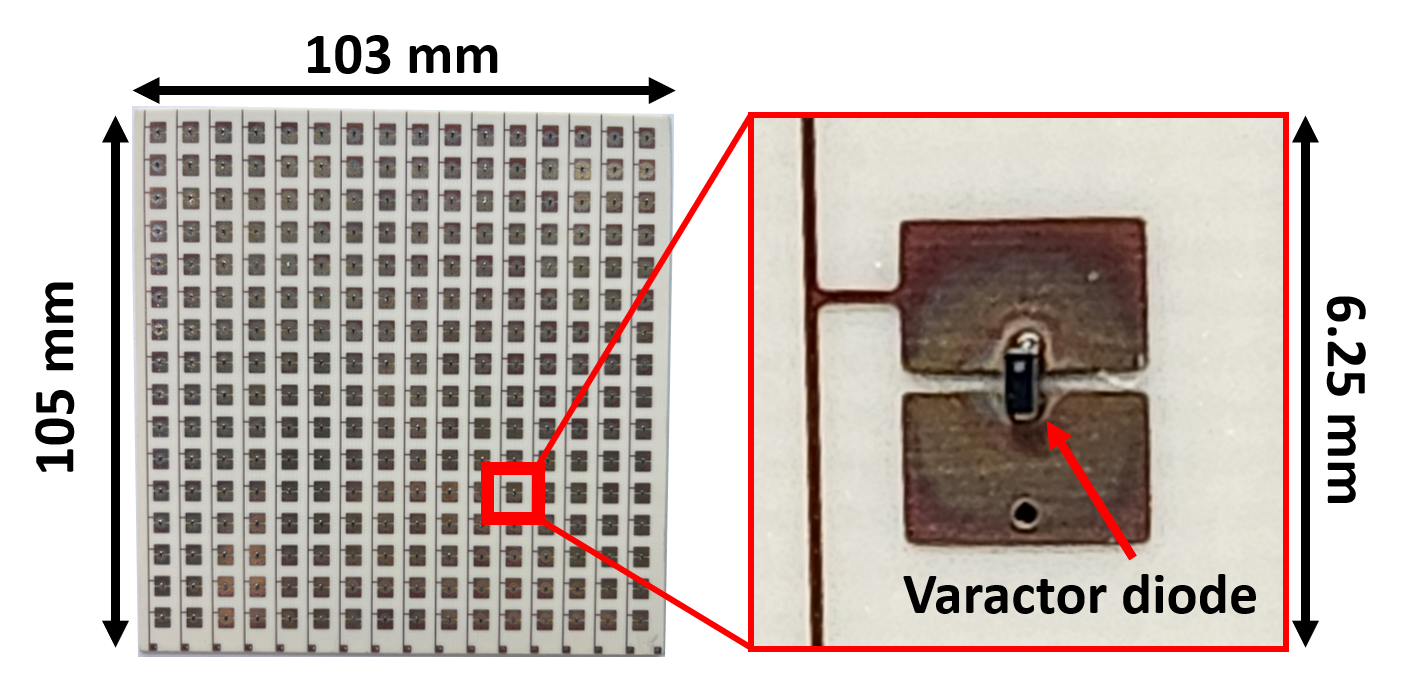}}
\vspace{-1mm}
\caption{Metasurface and a unit-cell.}\label{fig:hardware}
\vspace{-4mm}
\end{figure}

\vspace{-2mm}
\section{Prototype Implementation} \label{sec:impl}
\vspace{-1mm}

We fabricate a prototype of \ours and implement the real-time reconfiguration inside the NR deployment setup.

\vspace{-2mm}
\subsection{Hardware Configuration}
\vspace{-2mm}

\vspace{1mm} \noindent \textbf{Metasurface.} Figure~\ref{fig:hardware} shows the fabricated metasurface with the RO4003C substrate~\cite{Rogers} of 32mil thickness, consisting of 16x16 unit-cells. Each column is connected to a single bias line and configured by a GPIO interface of the MCU. The unit-cell is magnified in the red box, showing the MAVR-011020-1411~\cite{varactor_diode} varactor mounted on the center.

\vspace{2mm} \noindent \textbf{NBPU Rx.}
\ours employs a fully operational NBPU Rx aiming at a high SNR for reliable synchronization and decoding while minimizing the power consumption down to $\mu$W regime. 
Figure~\ref{fig:Rx_hardware} shows the NBPU Rx consisting of a filter, envelope detector, and two amplifiers. 
NBPU Rx first filters out signals in passband using the TFR915X\cite{TFR915X} Surface-Acoustic-Wave filter, whose center frequency is 915MHz with the 3dB-bandwidth 320KHz suitable for the NBPU's guardband operation of the NB spectrum. We note the passband filter is particularly important for the NBPU, as the envelope detector down-converts all signals into the same baseband.

NBPU Rx employs a MA4E2200~\cite{MA4E2200} Schottky diode to yield the envelope of the filtered NBPU signal. The diode is matched to the antenna with an L-matching network consisting of two inductors, 22nH in parallel and 48nH in series, reducing the impedance mismatch loss to -4dB. To boost SNR, NBPU Rx employs 2-stage $\mu$W-consuming MCP6G03~\cite{MCP6G03} amplifiers to boost the signal strength by +68dB in total, which can be controlled from +68dB to +40dB to avoid distortions at large signal strengths. We note the amplifier yields +14dB SNR on the typical operating scenario~\cite{38901}, suitable for tight synchronization which we demonstrated in Section~\ref{subsection:sync_performance}.

\vspace{2mm} \noindent \textbf{Power consumption.}
\ours consume the most energy during the tight synchronization and decoding of NBPU, up to $2.4m$W consisting of $655\mu$W for amplifier, $294\mu$W for ADC, $111\mu$W for the synchronization timer, and $1.36m$W for the synchronization and decoding logics. Leveraging duty-cycling, the power-hungry components are only active during a duration of NBPU which is much shorter than the entire 20ms period. 
This allows \ours to significantly reduce power consumption to $119.3\mu$W by 20.4x. 
Meanwhile, reconfigurations consume $1.67m$W from the synchronization timer and reconfiguration logic which converts beam patterns into GPIO states. 
This is also reduced with duty-cycling, down to $117\mu$W by 13.9x when 8 SSBs are assigned to \ours beam patterns. 
During the low-power mode, \ours consumes only $6.4\mu$W to hold configurations, which consists of $5.9\mu$W for the low-power mode standby current, $0.5\mu$W for a timer for reconfiguration timings, and $0.072\mu$W for power dissipated at unit-cells. \ours power consumption during the 20ms period sums up to a total $242.7\mu$W, which provides a 2.1-year lifetime from a single AA battery.
\vspace{-1mm}

\begin{figure*}[t!]
\centering
\includegraphics[width=2.1\columnwidth]{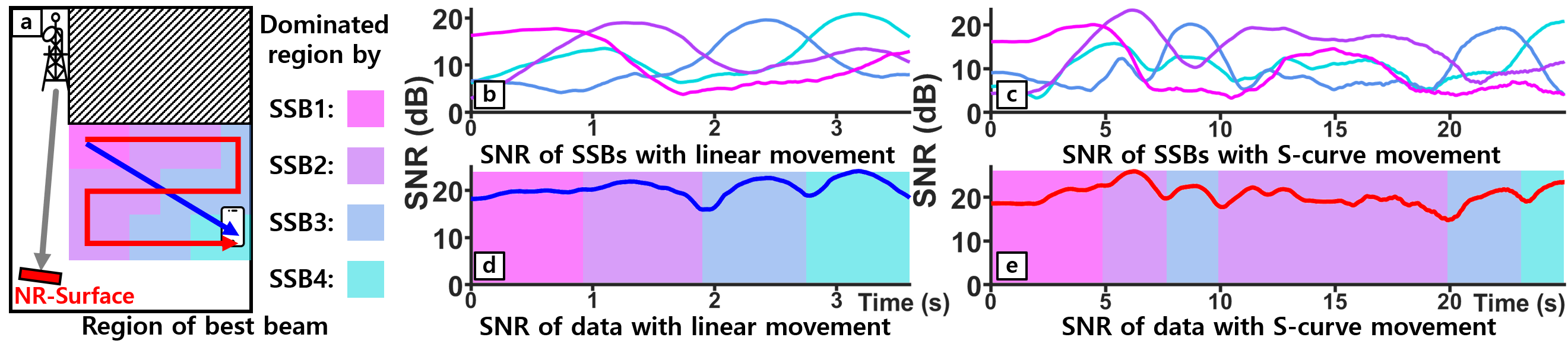}
\vspace{-6mm}
\caption{\ours's real-time reconfiguration performance under mobile UE scenario. (a) In the indoor experiment setting, \ours can increase coverage with multiple beams. Under various UE movements (e.g., linear \& S-curve), the best beam is changed by the location (b,c) and \ours reflects it in real time (d,e).} \label{fig:mobility}
\vspace{-3mm}
\end{figure*}

\vspace{-1mm}
\subsection{NR Evaluation Setup}
\vspace{-1mm}
We consider a general NR FR2 scenario. The BS is configured to transmit NB-IoT frames in FR1 containing the NBPU, and data packets in FR2 of large bandwidth.

\vspace{2mm} \noindent \textbf{FR2 transmissions.}
The BS employs USRP X300 series~\cite{X300} software-defined radio platform, transmitting NR-compatible signals generated from the MATLAB 5G Toolbox~\cite{matlab5gtoolbox}. The baseband signal is configured with 50MHz channel bandwidth, which consists of 64 SSBs for the beam sweeping followed by 15ms of data. The signal is continuously generated with a period of 20ms.
For FR2 communication, the baseband signal is upconverted by ADMV1013~\cite{ADMV1013} to mmWave ISM band centered at 24.125GHz for the prototype evaluation, with +5dBm transmit power. The FR2 signal is transmitted by the 24GHz microstrip patch array antenna~\cite{patch_antenna} with a fixed beam direction and +22dBi directivity.

The UE is also equipped with the same microstrip patch array antenna to receive the FR2 signal, which is downconverted by ADMV1014~\cite{ADMV1014}. The UE employs USRP X300 series to process the baseband frame of bandwidth 50MHz, running a custom GNURadio~\cite{GNURadio} block. The UE measures SNRs of SSBs in the beam sweeping, which is reported to the BS using the 905MHz spectrum within the ISM band. We note the RF characteristics of 900MHz ISM spectrum are also similar to NR low-band spectrum.

\vspace{2mm} \noindent \textbf{FR1 transmissions.}
The BS employs the same USRP running srsRAN~\cite{srsRAN}, an open-source software implementation of an O-RAN compliant base station. The BS is configured to 900MHz ISM band centered at 915MHz for the prototype evaluation, with 180KHz bandwidth for the NB-IoT signal. We provide NB-IoT payload bits such that the generated NB-IoT signal contains the NBPU. On the other hand, the BS receives the beam report from the UE and extracts the best SSB index, whose center frequency is configured to 905MHz within the ISM band. This SSB index is immediately reflected to the next NBPU transmission. For the transmission, the BS is equipped with a VERT900~\cite{VERT900} antenna of +3dBi gain.

\vspace{-1mm}
\section{Evaluation}
\vspace{-1mm}

In this section, we perform experiments under a dynamic environment to show that \ours is ready to be deployed in practice. Specifically, we demonstrate the practicality of \ours system by deploying \ours in mobile and rapidly changing environments such as human movement scenarios. We measure various performances required for practicality such as SNR gain, latency, coverage, robustness of NBPU, and power consumption. 

\vspace{-3mm}
\subsection{Dynamic Environment} \label{subsec:dynamic_environment_evaluation}
\vspace{-1mm}

\ours is ready for practical deployment in NR scenario, achieving both the high average SNR gain (max 22.2dB) with fast beam tracking performance ($<20$ms reconfiguration delay). In order to utilize \ours in real life, \ours must be able to quickly change the best beam direction in the dynamic environment that the mobile UE can experience. Since the environment around the handheld device is rapidly changed due to human movement, such as irregular movement or blockage by the human body, we evaluate the performance of \ours reconfiguration in both scenarios.

\begin{figure}[t]
\hspace{-0.1cm}
\centering
\includegraphics[width=1.0\columnwidth]{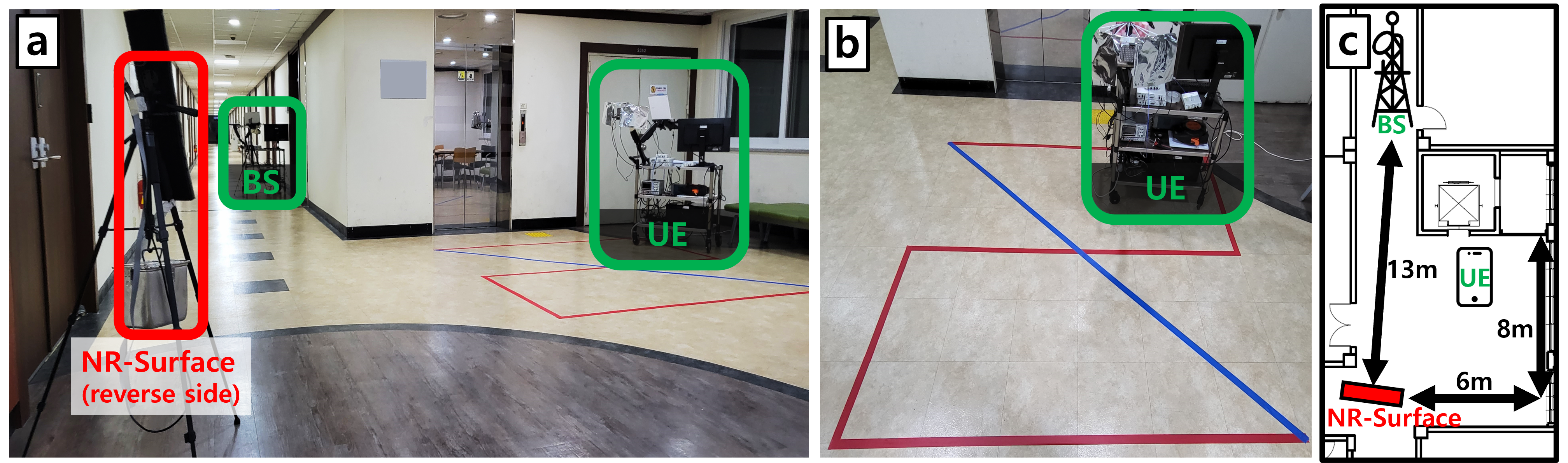}
\vspace{-0.5cm}
\caption{(a) Experiment setting for the mobile environment in the corridor and (b) the two tracks of mobile UE moved where the detailed scale is in (c).}\label{fig:exp_setting}
\vspace{-0.2cm}
\end{figure}

\begin{figure*}[h!]
\centering
    \includegraphics[width=2.1\columnwidth]{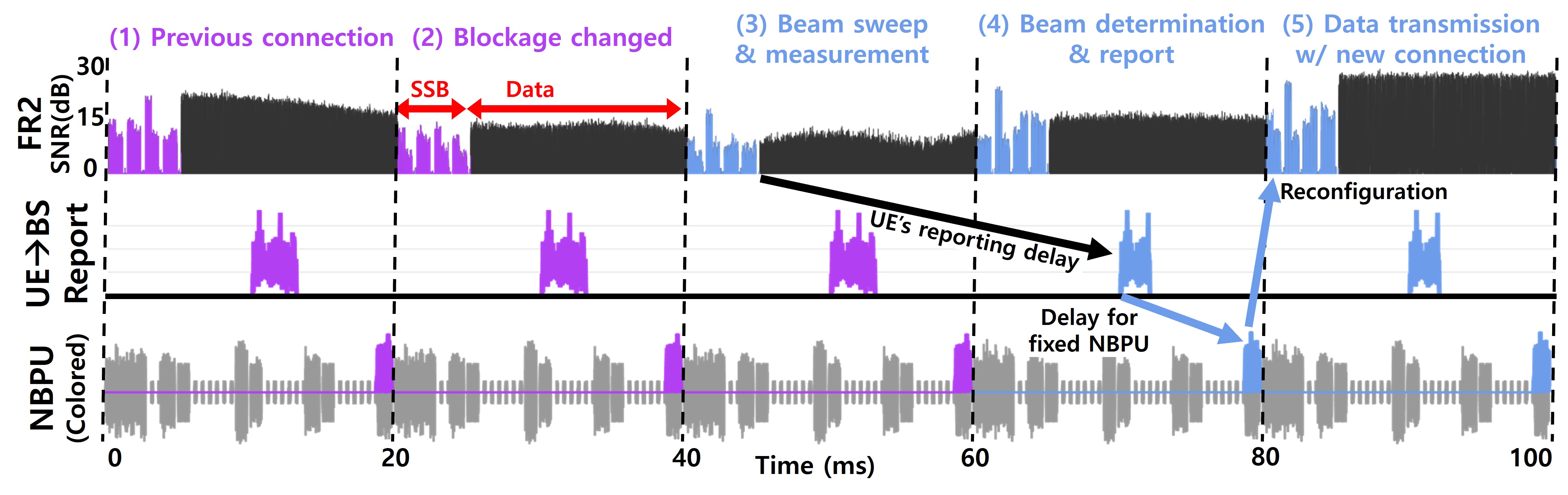}
    \vspace{-4mm}
    \caption{Real-world measurements of transmitted signals at the moment of blockage change. (1)-(3) Due to the dynamic blockage, the data SNR drops and the best SSB changes during the beam sweep. (4) The best SSB is determined and reported to the BS, which is also delivered to \ours within 20ms. (5) \ours is reconfigured to recover data SNR gain.}
    \label{fig:blockage}
\vspace{-1mm}
\end{figure*}

\begin{figure}[t]
\centering
\includegraphics[width=\columnwidth]{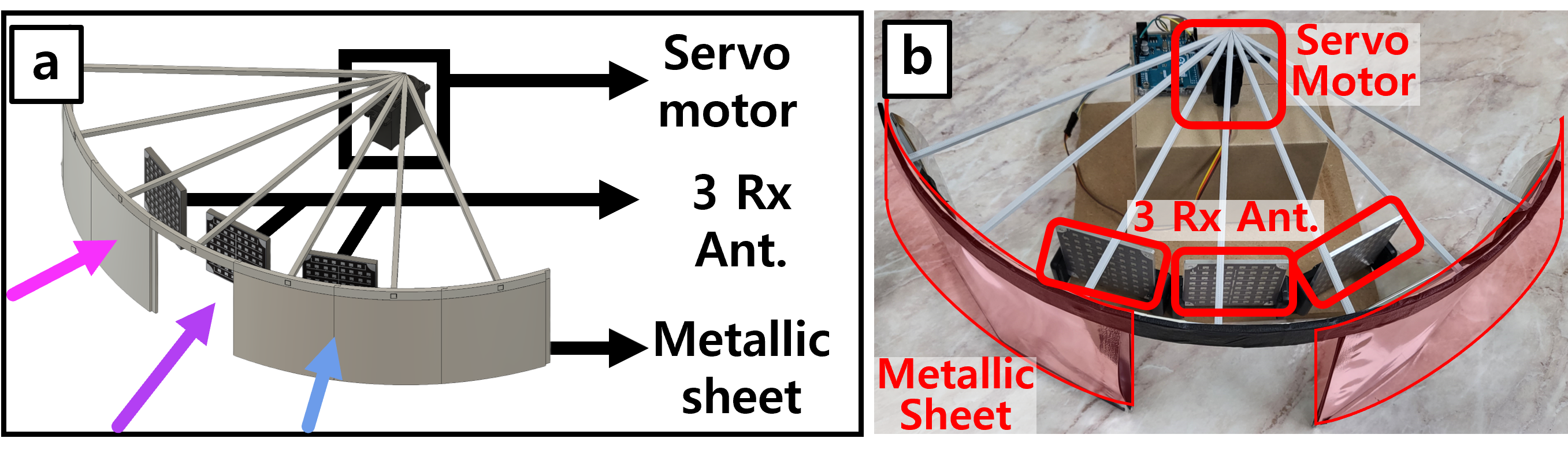}
\vspace{-7mm}
\caption{(a) outline of both blockage and UE design and (b) its implementation.}\label{fig:rx_schemetic}
\vspace{-5mm}
\end{figure}

\vspace{1mm} \noindent \textbf{SNR gain in mobile UE environment.} 
We evaluate the SNR gain in real-time reconfiguration of \ours with respect to the UE mobility. The evaluation setup is shown in Figure~\ref{fig:exp_setting}~(a) where the BS cannot support UE without \ours since the LoS path is blocked by the corridor. \ours is placed at the intersection to improve coverage of such UEs in the blind spot, which is tightly synchronized to the FR2 frame and sweeps beam configurations during the SS burst. The FR2 UE moves around the corridor following two paths in Figure~\ref{fig:exp_setting}~(b), namely "linear" in which UE moves at a typical human walking speed of 1m/s with constant distances to \ours, and "S-curve" which UE pass through multiple points with different distance to \ours. The detailed scale of our deployment is shown in Figure~\ref{fig:exp_setting}~(c).

As a reference, the region in Figure~\ref{fig:mobility}~(a) is colored depending on which beam configuration provides maximum SNR. Figure~\ref{fig:mobility}(b) and Figure~\ref{fig:mobility}(c) plot the dynamic SNR variation of 4 SSBs for the movement path "linear" and "S-curve", respectively, smoothed with window 10 over measurements taken once every 20ms. 
Specifically, UE with the linear movement passes through from SSB1 to SSB4 in sequence, and UE with the S-curve movement passes through in sequence as SSB1 $\rightarrow$ SSB2 $\rightarrow$ SSB3 $\rightarrow$ SSB2 $\rightarrow$ SSB3 $\rightarrow$ SSB4.
The SNR variation plots confirm that the best SSB changes as the UE passes through regions. 

Figure~\ref{fig:mobility}~(d) and Figure~\ref{fig:mobility}~(e) illustrate the SNR of data, measured every 20ms for the movement path "linear" and "S-shape" respectively.
In response to the change of optimal beam configuration when the UE crosses the virtual boundary, the UE reports the best SSB to NR BS every 20ms as a single fixed beam would quickly deteriorate. The colored backgrounds in these figures indicate the timing boundary of the best SSB changes. This is applied to \ours for real-time reconfiguration, such that the data SNR always follows the largest SNR among SSBs. \ours achieves high average SNR gain of 18.3dB and 20.3dB in each UE movement, respectively, without significant SNR drop at crossing boundaries.

\vspace{1mm} \noindent \textbf{SNR gain in blockage changing environment.}
Considering the human body can suddenly block the established beam direction, \ours should be able to rapidly update the best beam to cope with complex environments due to dynamically changing blockages.
In the NR beam management procedure, the UE experiences serious SNR drop when the LoS path is blocked by the dynamic environment. \ours deployment can recover the SNR drop by reconfiguring beam patterns in real time.

We setup the UE and blockage as in Figure~\ref{fig:rx_schemetic}(a)-(b), which has three microstrip patch antennas for Rx and movable metallic sheets via servo motor. Each three Rx antenna of UE is directed in different directions to check the best beam between multiple Rx beams, where each antenna is correlated to one Rx beam of digital beamforming.
For an accurate evaluation, the movement of blockage is mechanically controlled to minimize errors from manual manipulation. By rotating the servo motor accurately, the metallic sheets always block two antennas and allow only one antenna can receive the signal from \ours. Thus, after the blockage environment is changed, UE finds the best SSB during the beam sweep step among multiple Rx beamforming. Then, UE reports the best SSB index to BS, which reconfigures \ours in real time.

\begin{figure}[t]
\centering
\hspace{-1mm}
\includegraphics[width=\columnwidth]{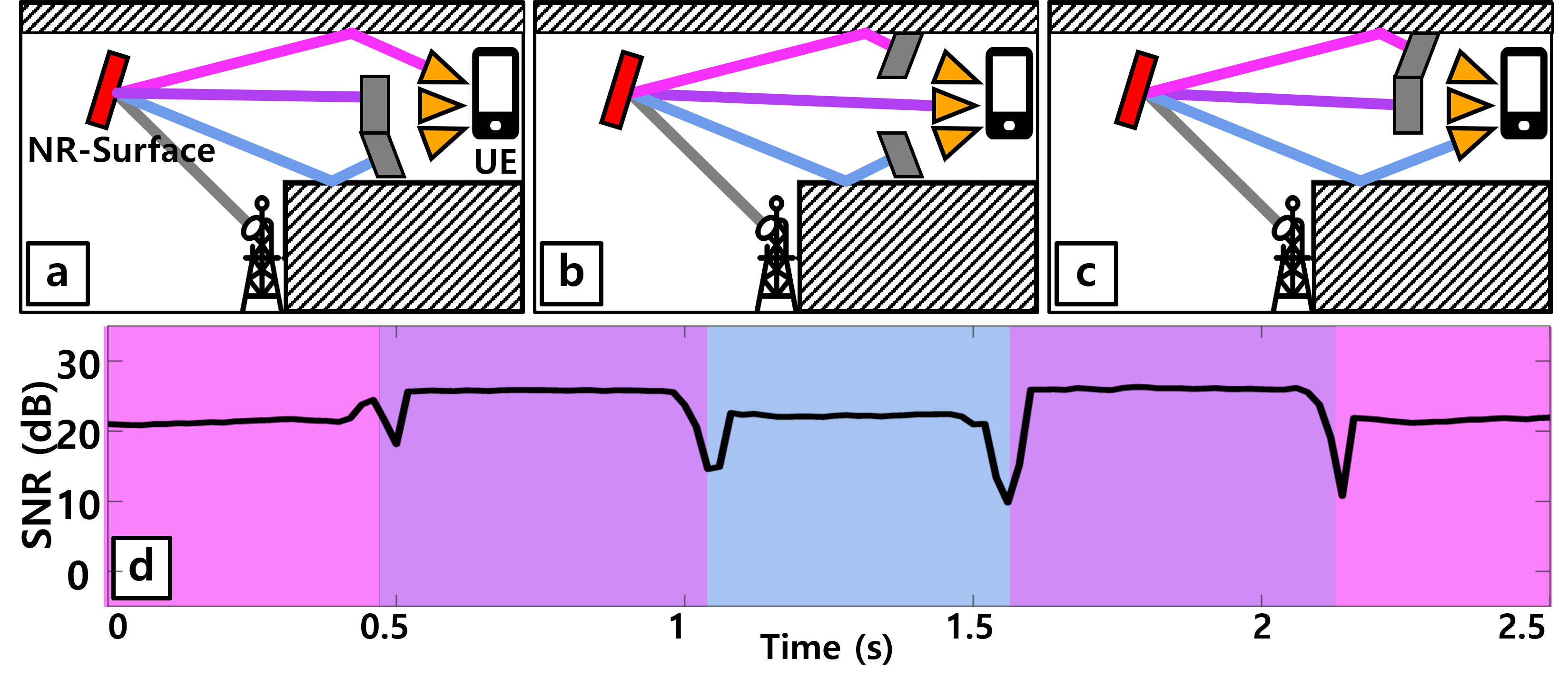}
\vspace{-5mm}
\caption{(a) - (c) show a scenario that only one antenna among 3 is not blocked. (d) shows the achieved SNR by the blockage environment change.}\label{fig:blockage_scenario}
\vspace{-3mm}
\end{figure}

Figure~\ref{fig:blockage_scenario}~(a) - (c) shows our evaluation setup, where the line-of-sight between BS and UE is blocked by the corridor. Each figure shows that only one antenna is allowed to receive a signal while the other is blocked by metallic sheets. We control the servo motor to change blockage quickly within 60ms, in (a) $\rightarrow$ (b) $\rightarrow$ (c) $\rightarrow$ (b) $\rightarrow$ (a) sequence where the servo motor is paused for 400ms in each state. Figure~\ref{fig:blockage_scenario}(d) plots the SNR gain of data every 20ms, where colored regions show the beam index in NBPU. SNR drops slightly between regions due to fast servo movements, and a stable SNR at paused durations. \ours can establish a high SNR path with 22.2dB average SNR by fast recovery.

\vspace{2mm} \noindent \textbf{Time domain analysis for beam establishment.} 
To get a closer look at what control sequence actually exists at the moment of blockage change, Figure~\ref{fig:blockage} shows what the received signals by UE, BS, and \ours look like by zooming in on the moment of blockage change from Figure~\ref{fig:blockage_scenario}~(c) to (b).
The FR2 SNR plot demonstrates the strongest SSB is changed during (1)-(3) due to the new blockage setting. Thus, the SNR gain of the data at (2) is lower than the previous data at (1). Then, during the next beam sweep step at (3), UE newly determines the best SSB among 3 Rx antennas. After UE's beam determination, UE reports the best SSB index to BS via the FR1 channel in (4), where a few milliseconds (25 ms - 29 ms in our implementation) delay occurs due to UE's processing time and allocated UL time. Note that this UE-driven delay can be shorter when UE has a powerful signal processing unit or BS allocates resources more frequently to UE.
After BS gets a report from UE, BS transmits NBPU to control, shown as the blue solid line in (1). The fixed time of NBPU causes a max 20 ms delay, which is short enough to recover the blocked path by human movement. Finally, \ours is reconfigured to the new SSB index in (5).

\begin{figure}[h!]
\centering
\hspace{-2mm}
\includegraphics[width=\columnwidth]{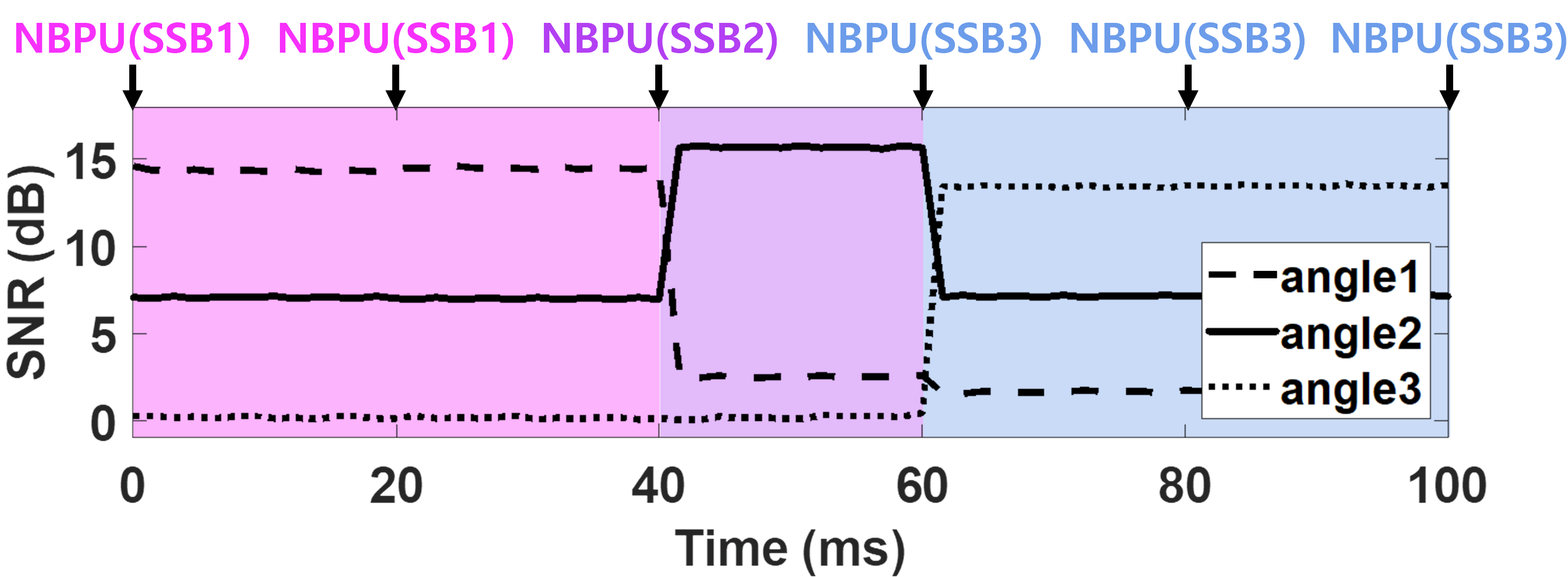}
\vspace{-2mm}
\caption{SNR of three beam patterns when \ours is reconfigured in real-time. \ours continuously receives NBPU and reconfigures beam patterns every 20ms.}\label{fig:fast_servo}
\end{figure}

\begin{figure}[h!]
\centering
\includegraphics[width=\columnwidth]{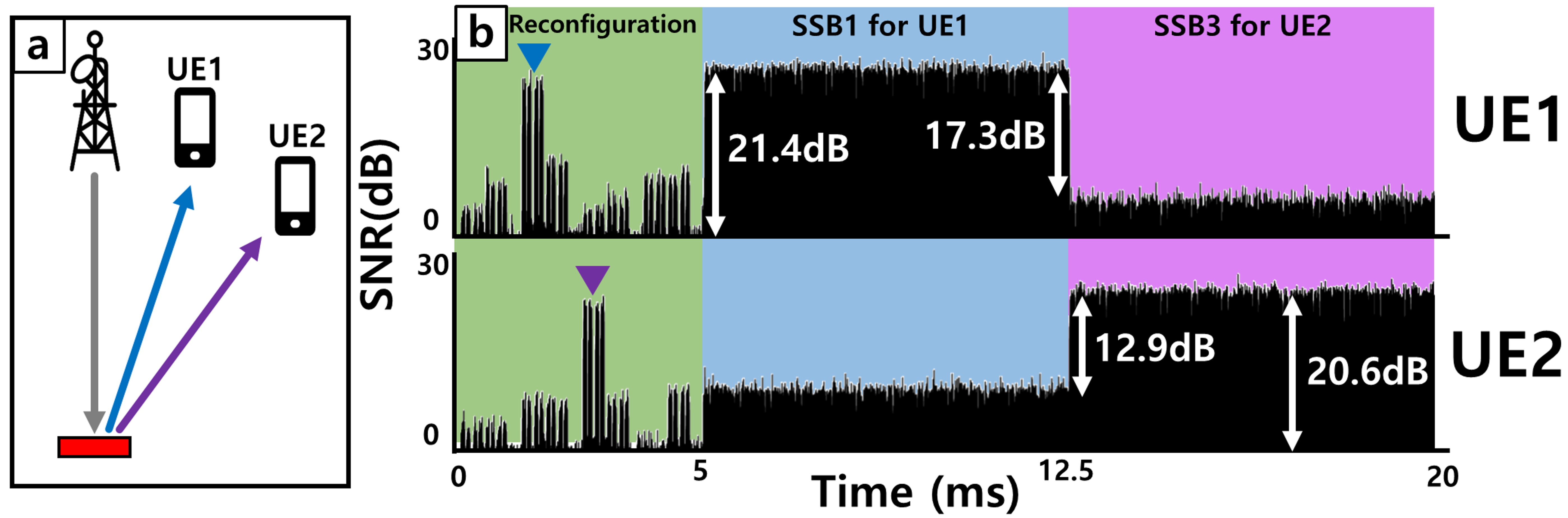}
\vspace{-5mm}
\caption{(a) \ours operation scenario with multiple UEs (b) SNR fluctuation of the received FR2 per each UE.}\label{fig:multi_UE}
\vspace{-2mm}
\end{figure}

\vspace{2mm}
\noindent \textbf{Real-time configuration analysis.}
\ours real-time reconfiguration can support highly dynamic environments where the best beam pattern changes every 20ms period of the NR beam management procedure. We employ the same metal-sheet blockage and a fast servo that changes blockage within 20ms, such that the UE reports different beam patterns in each NR beam management procedure. This is also delivered to \ours via NBPU every 20ms. Figure~\ref{fig:fast_servo} plots the SNR of three antennas placed at the main lobe of beam patterns, which demonstrates \ours being continuously reconfigured every 20ms to support dynamic environments in real-time.

\vspace{-2mm}
\subsection{Multi UE Operation}
To demonstrate multi-UE operation, we place two UEs at different locations such that each UE can select different SSBs for its data transmission (Figure~\ref{fig:multi_UE}(a)). UE1 and UE2 are scheduled to receive data from 5ms to 12.5ms, and 12.5ms to 20ms, respectively. \ours receives the schedule info along with SSB indices.
Figure~\ref{fig:multi_UE}(b) illustrates the SNR of the received FR2 signal at each UE. During the beam sweeping, UEs report different SSBs as their best SSB index, which are SSB1 and SSB3 for UE1 and UE2 (triangles in Figure~\ref{fig:multi_UE}), respectively. 
\ours is reconfigured to SSB1 at 5ms for the UE1, such that the UE1 achieves +21.4dB SNR during its data transmission. Then \ours is reconfigured at 12.5ms such that the UE2 achieves +20.6dB SNR from 12.5ms to 20ms. These SNRs are +17.3dB and +12.9dB better when compared to \ours is reconfigured to only a single beam pattern of the other UE, for UE1 and UE2, respectively.

\begin{figure}[h]
\centering
\includegraphics[width=0.9\columnwidth]{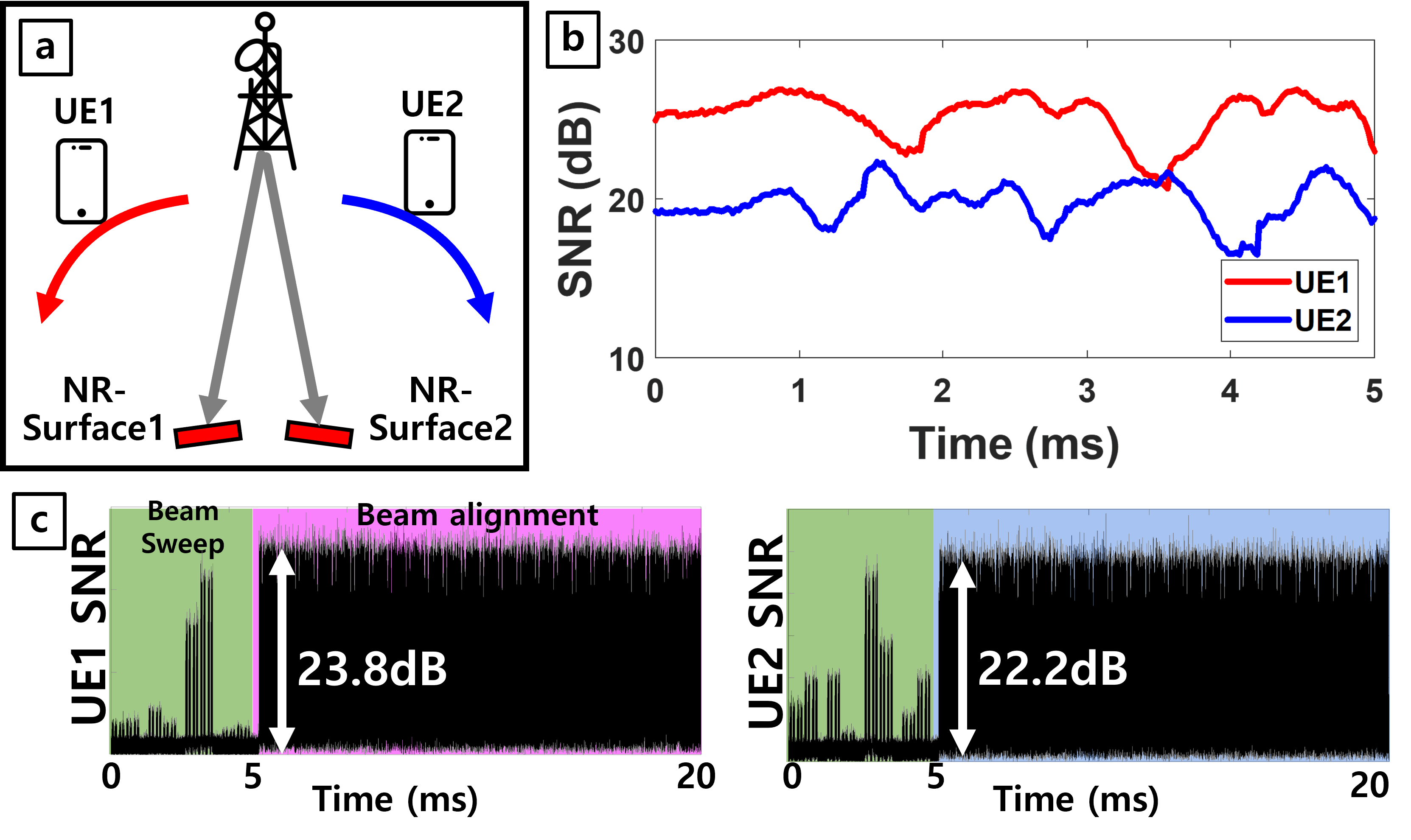}
\vspace{-1mm}
\caption{(a) Multiple \ours scenario with multiple moving UEs (b) SNR fluctuation of the received FR2 per each UE (c) Received SNR of each UE, while both users moved during 5 sec.}\label{fig:multi_MTS}
\vspace{-4mm}
\end{figure}

\subsection{Multi \ours} \label{subsec:multi_eval}

\ours supports deployment scenarios where one BS controls multiple \ours. 
Figure~\ref{fig:multi_MTS}(a) illustrates the evaluation setup where two \ours are placed and two UEs move around \ours. 
Note the NR BS reconfigures its beam towards multiple \ours, where the evaluation setup comprises two fixed antennas at the BS for demonstration purposes. 
Two NBPUs are placed in the same NB-IoT guardband at different subframes, with different IDs in the reconfiguration info to control two \ours separately. 
Figure~\ref{fig:multi_MTS}(b) plots the data SNR of each UE, showing each \ours is concurrently reconfigured in real-time to track each UE, respectively, with an average $22.5$dB SNR gain. Figure~\ref{fig:multi_MTS}(c) plots the example 20ms period during the movement, showing each UE experiences different SSB as their best beam pattern. This is separately reflected to two \ours such that both UE achieve high data SNR. The evaluation demonstrates the capability of multiple \ours deployment in complex environments, providing a large number of SSBs to cover multiple blind spots.
\vspace{-2mm}

\begin{figure}[h!]
\centering
\includegraphics[width=0.9\columnwidth]{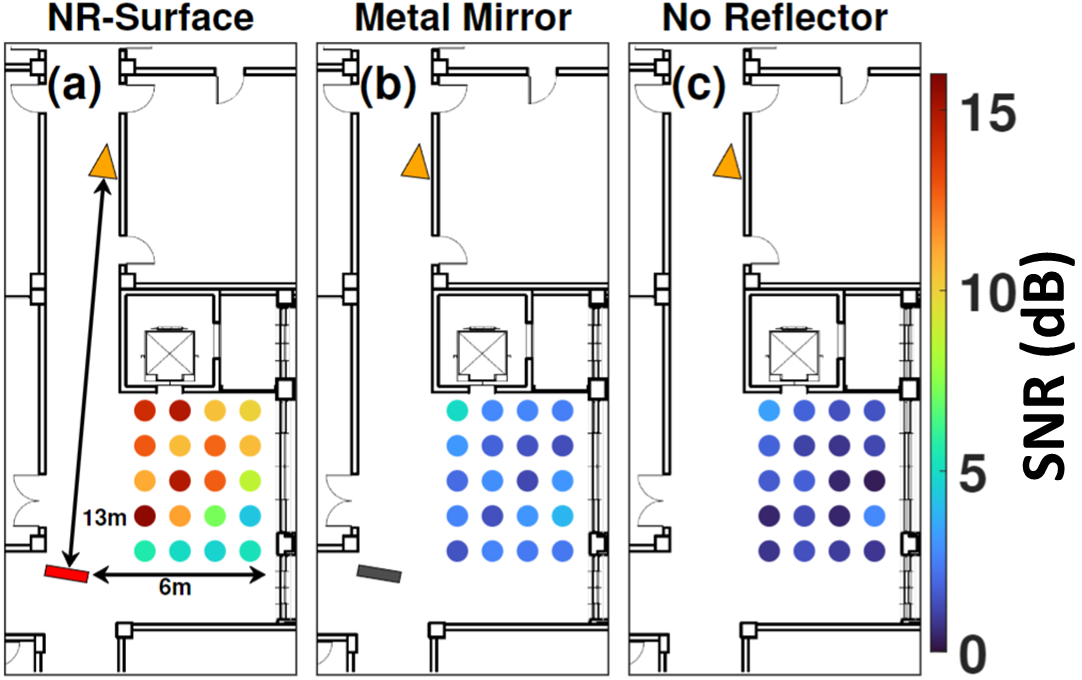}
\vspace{-2mm}
\caption{Comparison of SNR at various Rx locations (dots) between (a) \ours placed at red rectangle (b) metal mirror of the same size with \ours (c) no reflector placed. The Tx is placed at orange triangle steered towards \ours.}\label{fig:extensive}
\vspace{-2mm}
\end{figure}

\subsection{Coverage Extension}

We evaluate SNR improvements of \ours in various UE locations, compared to a naive metal reflector and no reflector being placed.
Figure \ref{fig:extensive}(a) illustrates the measurement setup, a typical around-the-corner scenario with a strong blockage between Tx and Rx. Tx to reflector distance is fixed at 13m, and the transmit power is configured to 10dB less than previous experiments to emulate more practical attenuation. The SNR is measured at 4x5 UE locations on a rectangular grid of 1m spacing, with a maximum Rx to reflector distance is 7.3m. \ours and metal mirror are placed at the intersection facing the Tx, where \ours is reconfigured to steer the best beam possible while the metal mirror is fixed. Figure \ref{fig:extensive}(a)-(c) show \ours can achieve on average 7.6dB gain than the SNR with a metal mirror, and 8.8dB gain than the SNR without reflectors. Note the SNR improvement diminishes at UE locations whose steering angle exceeds $70\degree$ from the normal direction, showing \ours can support wide range of steering angles up to $140\degree$ in practice.

\begin{figure}[!h]
\begin{minipage}[t]{0.48\linewidth}
    \includegraphics[width=\linewidth]{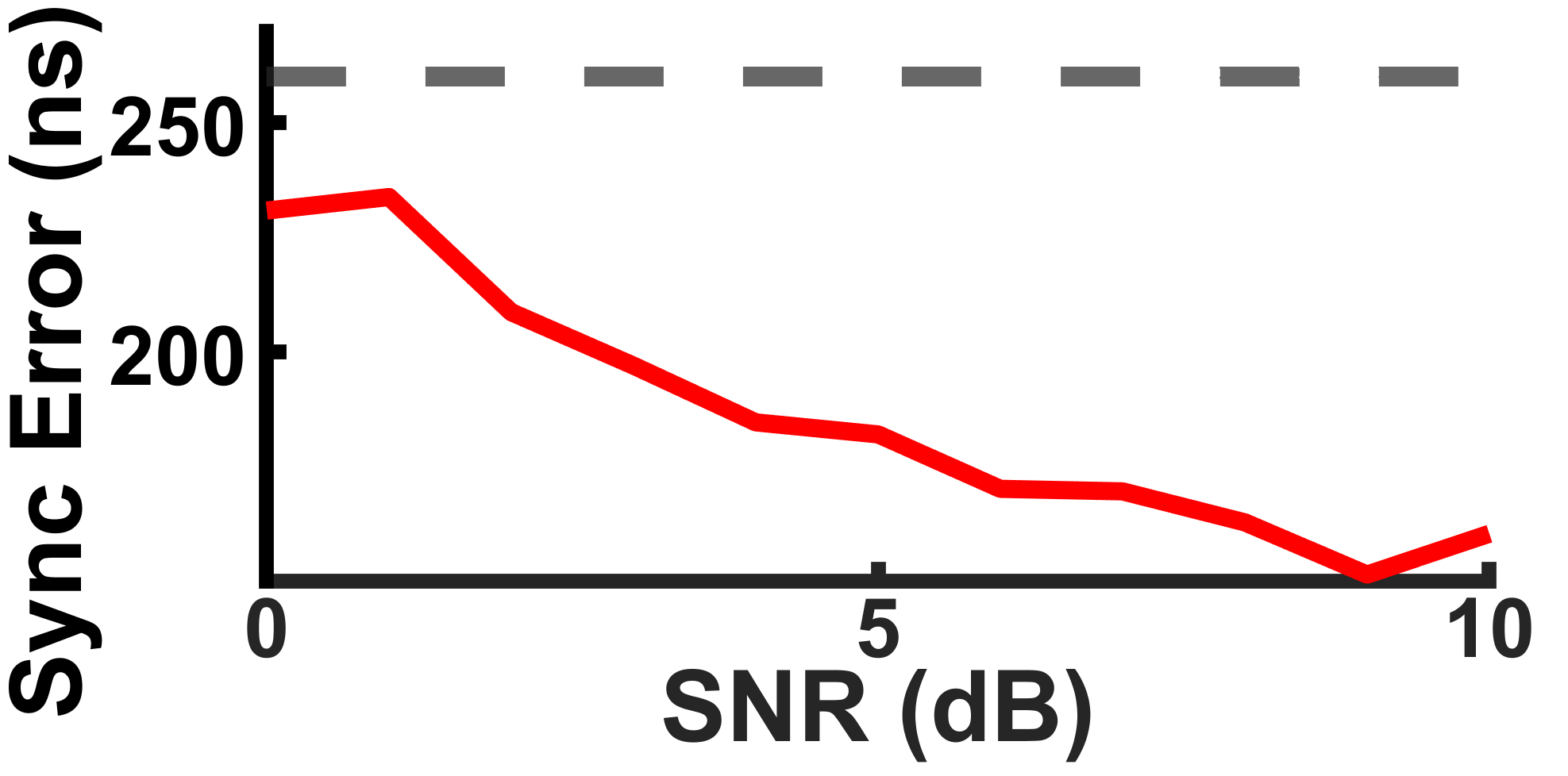}
    \caption{NBPU Synchronization error versus SNR.}
    \label{fig:sync}
\end{minipage}%
    \hfill%
\begin{minipage}[t]{0.48\linewidth}
    \includegraphics[width=\linewidth]{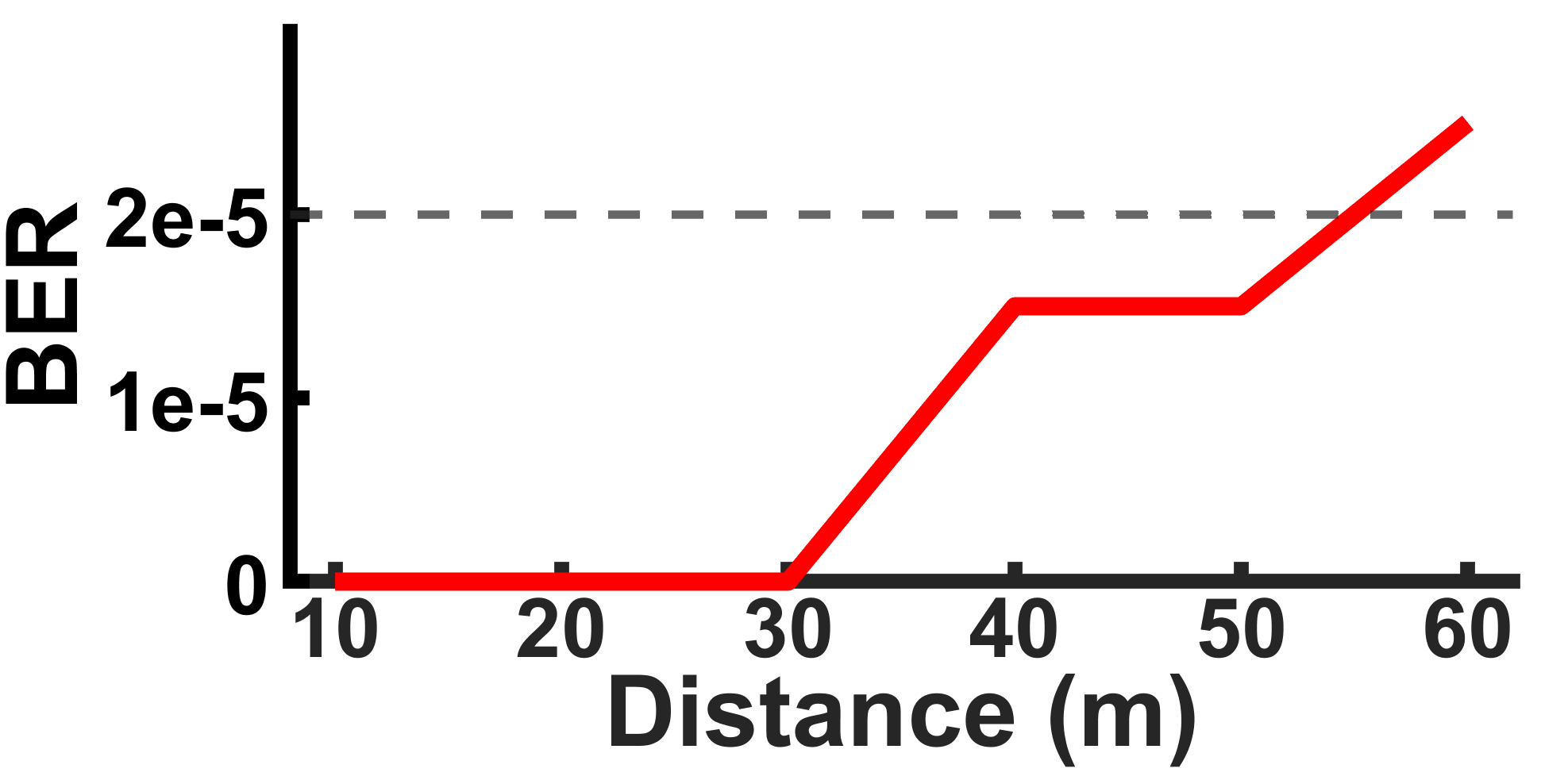}
    \caption{\ours BER versus distance to the BS.}
    \label{fig:ber}
\end{minipage} 
\vspace{-4mm}
\end{figure}

\vspace{-1mm}
\subsection{\texorpdfstring{$\mu$}{\textmu}W Synchronization Accuracy}\label{subsection:sync_performance}
\vspace{-1mm}

We evaluate \ours synchronization accuracy which is necessary to ensure compliance with NR standards. Figure~\ref{fig:sync} plots the average synchronization error with respect to the SNR over 400 NBPUs, measured as the difference between the boundary of the NBPU transmission and the rising edge of the GPIO interface in \ours. Specifically, we connect a high sampling rate ($\sim1Gsps$) oscilloscope to both \ours Rx and GPIO interface. 
\ours successfully achieves $234$ns synchronization error at low SNR of $+0$dB, which is much lower than SNR in typical operation scenarios~\cite{38901}. From this, reconfigurations satisfy the NR requirements of $<260$ns including $<10$ns reconfiguration delay, even under the typical clock drift $<20$ns~\cite{Diodes_white, quartz_dsheet} quartz crystal occurring every 20ms period. \ours achieves this high accuracy in low-power by the equivalent time sampling, which mimics 3.84MHz ADC with a low-end 14KHz ADC.

\vspace{-1mm}
\subsection{NBPU BER Measurements}
\vspace{-1mm}
We evaluate \ours demodulation performance and corresponding reliability of \ours reconfigurations. Figure~\ref{fig:ber} plots the average BER of \ours with respect to the distance to the NR BS, averaged over 200,000 symbols. Specifically, NR BS is placed in the indoor corridor and transmits NBPU at +16dBm, the maximum transmit power of NB-IoT in small cell BS. \ours is equipped with the +11dBi Yagi antenna towards the BS, assuming antenna direction is known during the installation. \ours achieves $<2\times10^{-5}$ BER up to 50m distance, which corresponds to $99.99$\% satisfying high data-rate applications by NR KPIs~\cite{22261}, assuming 5 bits are used for reconfiguration info. \ours achieves this in low-power by asymmetric NBPU OOK symbol design maximizing SNR while reducing bandwidth for low-power \ours Rx.

\begin{figure}[h!]
\centering
\subfigure[Coexistence of NBPU and in-band FR1.]{\includegraphics[width=0.56\columnwidth]{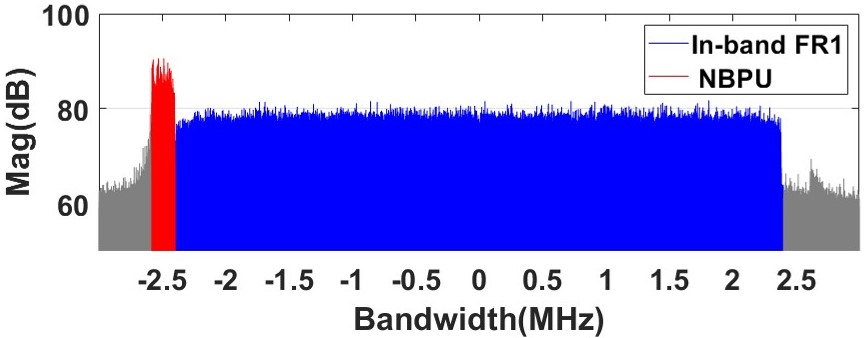}}
\hspace{0.5mm}
\subfigure[BER versus SINR in NR guardband]{\includegraphics[width=0.41\columnwidth]{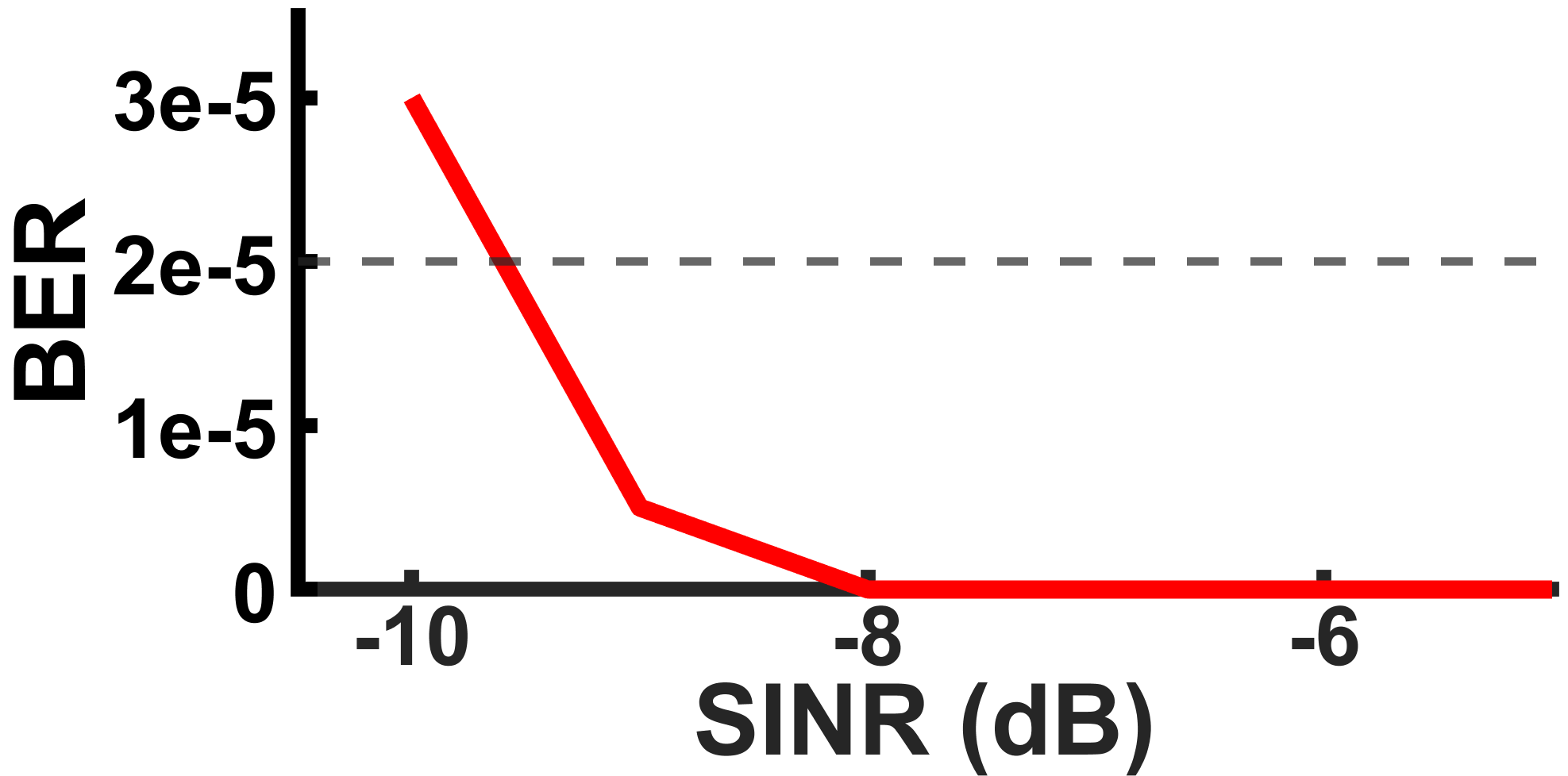}}
\vspace{-2mm}
\caption{Evaluation setup and BER performance against NR interference.} \label{fig:Coexist_freq}
\vspace{-4mm}
\end{figure}

\vspace{-1mm}
\subsection{Interference Robustness}
\vspace{-1mm}

We evaluate \ours robustness against the NR interference. NBPU is implemented on top of NB-IoT, which is placed at guardbands of the NR spectrum. Figure~\ref{fig:Coexist_freq}(a) plots the fully utilized 5MHz NR spectrum consisting of heavy NR traffic (blue) and NBPU (red) in the guardband, placed with only a 70KHz gap between the NR traffic. Note the NB-IoT in the guardband is allowed to boost transmit power by +6dB compared to the NR spectrum. Figure~\ref{fig:Coexist_freq}(b) plots the BER with respect to the SINR. The SINR drops down to $-8$dB when the NR spectrum is fully utilized. However, \ours successfully achieves $<2\times10^{-5}$ BER even at maximum interference (up to $-10$dB). \ours achieves this by a narrow passband filter of $320$KHz bandwidth to suppress the interference by $-18$dB, such that \ours still achieves low BER and high reliability in practical NR deployments. 

\begin{figure}[h!]
\centering
\subfigure[Simulated beam pattern]{\includegraphics[width=0.48\columnwidth]{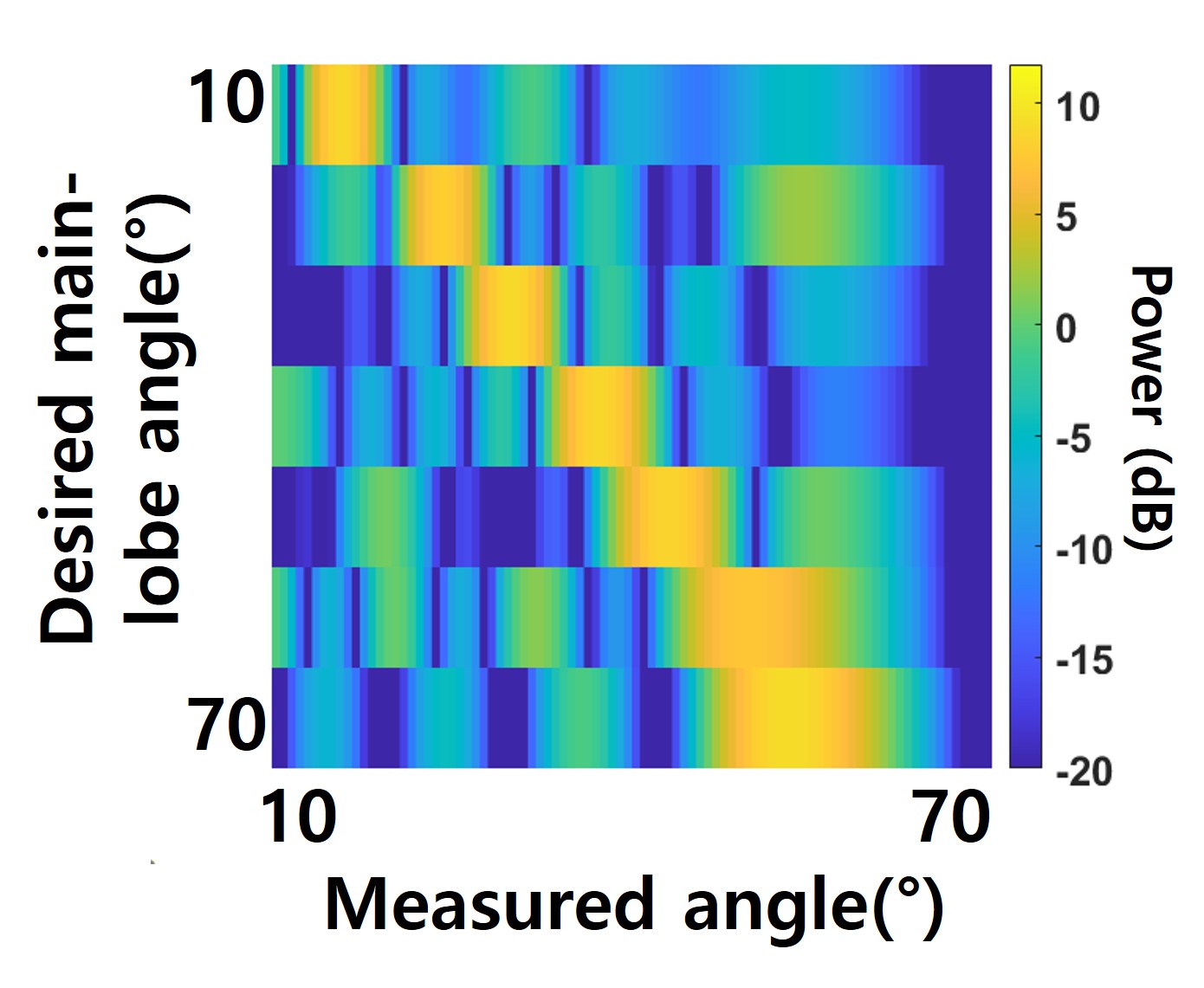}\label{fig:beam_simul}}
\subfigure[Measured beam pattern]{\includegraphics[width=0.48\columnwidth]{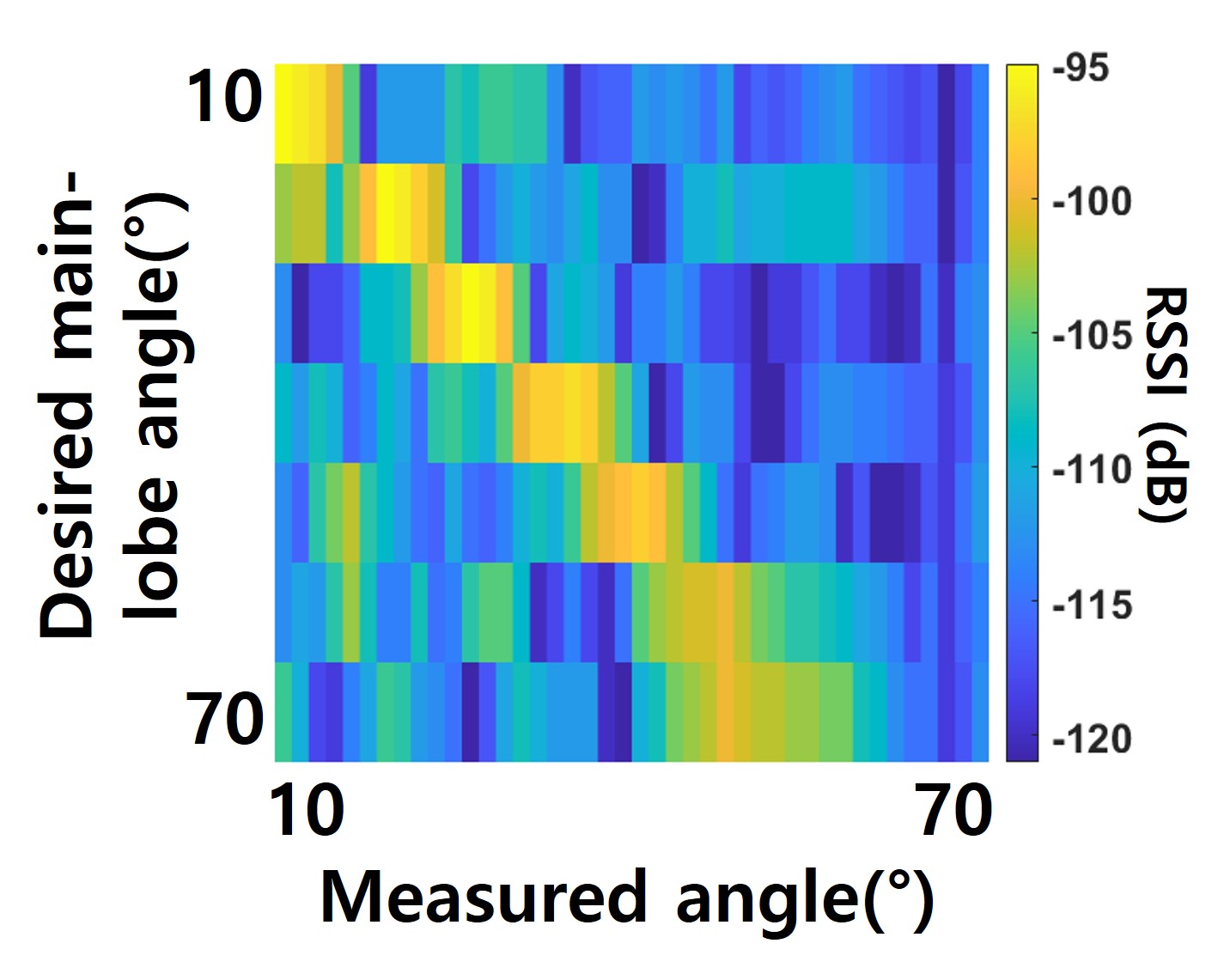}\label{fig:beam_eval}}
\vspace{-4mm}
\caption{Beam patterns with desired main-lobe angles from $10\degree$ to $70\degree$.} \label{fig:beam_pattern}
\vspace{-4mm}
\end{figure}

\subsection{Beam Pattern Validation}
\vspace{-1mm}
Figure~\ref{fig:beam_pattern}(a) and (b) illustrate the simulated and measured beam patterns generated from \ours using 1-bit GPIO outputs, respectively. Power simulation and measurement results are arranged such that each row represents a single beam pattern desired to be steered towards $+10\degree$ to $+70\degree$. Note that the measurement resolution is $2\degree$. The power measurement matrix has a similar appearance to the simulation, having a strong diagonal demonstrating each beam pattern is successfully steered to the desired angle. \ours successfully covers up to $140\degree$ from $-70\degree$ to $+70\degree$ using symmetric patterns, where side-lobes of the beam pattern are suppressed. 

\begin{figure}[h!]
\vspace{-1mm}
\centering
\includegraphics[width=\columnwidth]{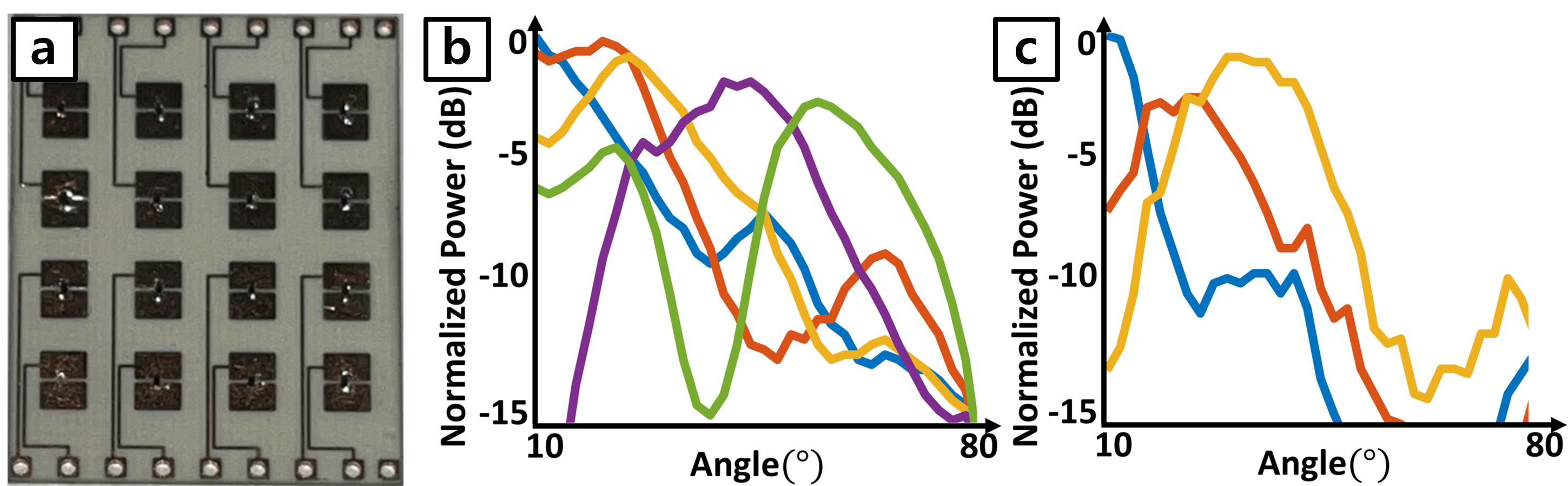}
\vspace{-4mm}
\caption{(a) Fabricated 3D metasurface and (b) measured 3D Beam pattern of \ours steered towards different azimuth angles and (c) different elevation angles direction (smoothed with window=$10\degree$).} \label{fig:beam_3D}
\vspace{-7mm}
\end{figure}

\subsection{Extension to 3D Beamforming} \label{sec:3D_beampattern}
\vspace{-1mm}
To demonstrate the 3D beamforming capability, we fabricate a 4x4 metasurface whose unit-cells are separately controlled. Figure~\ref{fig:beam_3D}(a) shows the fabricated metasurface, whose bias lines are separately connected to different bias terminals. Figure~\ref{fig:beam_3D}(b),(c) illustrates measured beam patterns steered toward different azimuth angles and different elevation angles. We note the valid beam patterns with strong main-lobe are drawn in Figure~\ref{fig:beam_3D}(b),(c).
\ours successfully steers beam towards azimuth angles from $+10\degree$ to $+60\degree$, covering $120\degree$ from $-60\degree$ to $+60\degree$ using symmetric patterns.
With the same 3D metasurface, \ours can also steer beams toward elevation directions from $+10\degree$ to $+40\degree$, covering $90\degree$ from $-45\degree$ to $+45\degree$ using symmetric patterns. We note the slightly narrower elevation coverage can be compensated with multiple \ours deployments.
\ours successfully steers the beam to different directions in 3D, which can be further extended to narrower and stronger beam patterns when implemented with larger metasurfaces.

\vspace{-3mm}
\section{Related Works}
\vspace{-2.5mm}
mmWave communication is considered to be the future of mobile communications, on which extensive research has been conducted~\cite{ghoshal2022depth, mazaheri2019millimeter, ahmad2019learning, eid2019scalable, jog2020millimeter}. However, mmWave communication suffers from high attenuation, such that antenna arrays for effective beamforming are necessary~\cite{mazaheri2019low, ghasempour2018multi, haider2018listeer, hassanieh2018fast, jog2019many, madani2021practical, hawkeye}.
In efforts to increase the coverage and reliability of mmWave systems, approaches such as dense deployment of base stations (BS)~\cite{NRsmallcellFumihide,NRsmallcell}, dual-band wireless network architecture with sub-6GHz~\cite{yan2022kf}, relays~\cite{ntontin2019relay, naqvi2021achieving, bi2022proximity}, and distributed antennas~\cite{mahboob2019fiber,sung20195g,moerman2022beyond,naqvi20215g} has been proposed. While simple and straightforward, these solutions require installing a large body of costly transceivers, hindering pervasive applicability and widespread of mmWave technology.

Metasurface~\cite{zhang2023acoustic, zhang2021rss} has emerged as an economic solution for aiding mmWave beamforming for reliable communication with extended coverage. Specifically, \ours cost (i.e., $\sim \$600$) is much cheaper than COTS repeater solutions pricing $\sim\$3000$~\cite{movandi}. Furthermore, \ours still has a similar cost to the existing mmWave metasurface with the same or smaller size~\cite{twoDimMTS, GrosMTS, WolffMTS} which cost $\$500-\$800$.
They support dynamic channel~\cite{cho2021mmwall, cho2022towards}, MIMO~\cite{zhang2021beyond}, or even utilize low-cost 3D printing~\cite{10.1145/3495243.3517024}.
A recent metasurface, mmWall~\cite{cho2023mmwall}, integrates link layer discovery protocol of mmWave networks into the design. 
Nevertheless, real-time, NR-compliant MAC layer design still remains a challenge.
In the given context, \ours prioritizes seamless integration with NR standards, presenting distinct challenges such as achieving tight synchronization in low-power consumption.

Another line of research~\cite{mazaheri2021mmtag,lynch2022ultra, lynch20215g, lynch20235g, eid2020mm, soltanaghaei2021tagfi, kludze2023leakyscatter} focuses low power communication for sustainable deployment. Recent works~\cite{lu2023millimeter, kapetanovic2019glaze} are capable of demodulating network symbols with mW level power consumption. \ours uniquely implements the low power Rx for metasurface control, to operate with 247 $\mu W$ end-to-end for practical deployment.
\vspace{2mm}

\vspace{-5mm}
\section{Conclusion}
\vspace{-2mm}
We present \ours, an NR-compliant metasurface reconfigured in real-time within the NR standard. \ours operates at $\mu$W-regime achieving a 2.1-year lifetime on an AA battery, consisting of the metasurface and NR-compliant Rx, which are duty-cycled leveraging the unique features of the NR standards. \ours end-to-end system is implemented on srsRAN and evaluated on diverse scenarios from dynamic environments, multiple UEs to 3D beamforming achieving up to 22.2dB gain.

\vspace{1.5mm}

    \vspace{-5mm}

    \section*{Acknowledgements}
    \vspace{-1.5mm}
    We thank our shepherd, Prof. Haitham Hassanieh, and the anonymous reviewers for their constructive comments. This research was supported in part by the MSIT(Ministry of Science and ICT), Korea, under the ITRC(Information Technology Research Center) support program(IITP-2024-2020-0-01787) supervised by the IITP(Institute of Information \& Communications Technology Planning \& Evaluation), the IITP grant funded by the Korea government (MSIT) (No.2021-000269, Development of sub-THz band wireless transmission and access core technology for 6G Tbps data rate, \%33), and the National Research Foundation of Korea(NRF) grant funded by the Korea government (MSIT) (No. RS-2023-00213816). We would like to acknowledge the technical support from ANSYS Korea.

    \bibliographystyle{plain}
    \bibliography{ref}

\newpage

\begin{figure}[H]
\centering
\subfigure[NB-IoT downlink frame structure in guardband mode] {\includegraphics[width=0.95\columnwidth]{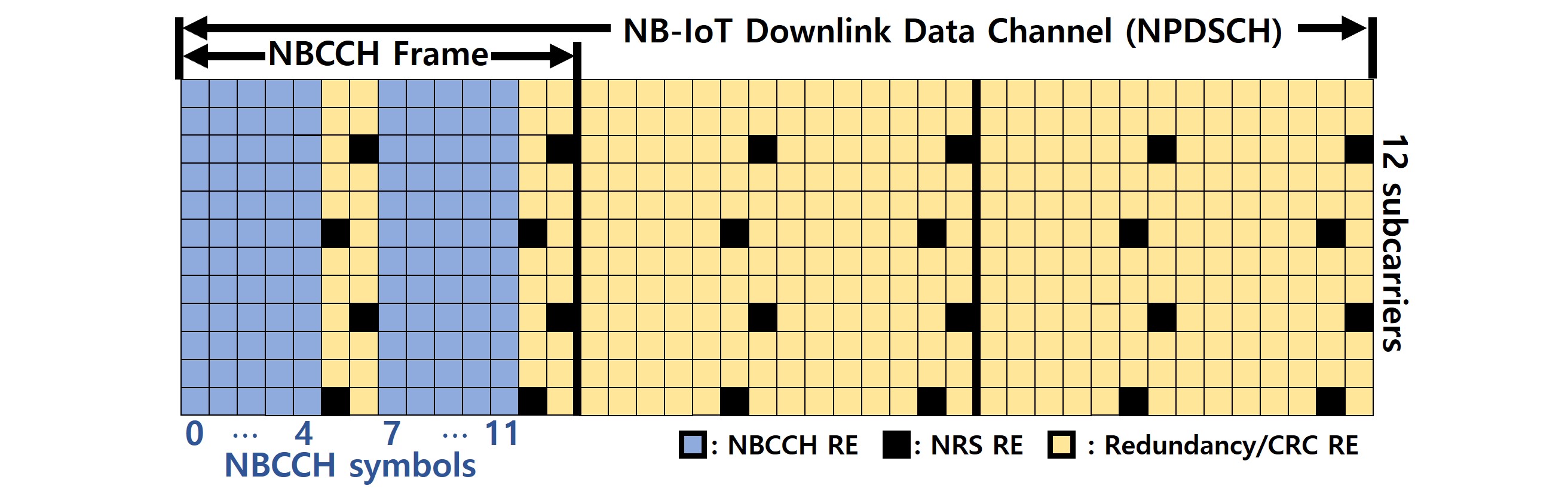}
\label{fig:NBIoT_structure}}
\vspace{-0.1cm}
\subfigure[NB-IoT PHY block diagram] {\includegraphics[width=0.95\columnwidth]{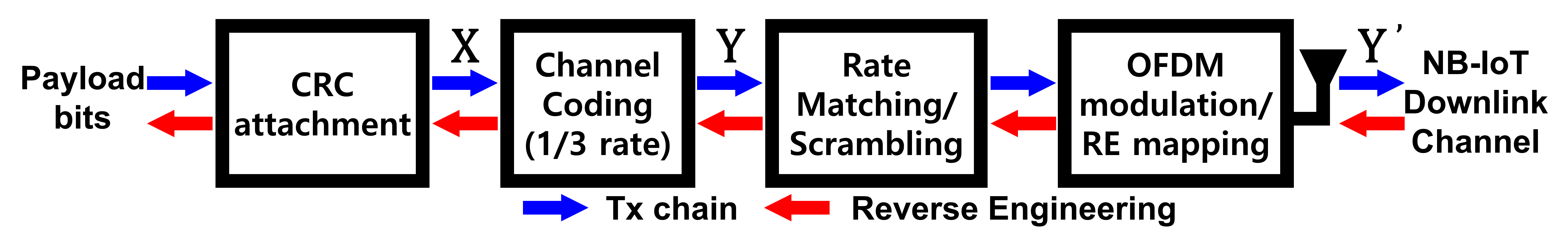}
\label{fig:reverse_sequence}}
\vspace{-0.3cm}
\caption{Emulating NBPU with NB-IoT}
\label{fig:reverse_engineering}
\vspace{-0.2cm}
\end{figure} 

\vspace{-5mm}
\section*{Appendix A. Emulating NBPU} \label{sec:appendix_a}

\vspace{-3mm}
As shown in Figure~\ref{fig:reverse_engineering}, NBPU is embedded using the emulation technique~\cite{li2017webee, chai2016lte, jeong2020sdr} in the NB-IoT downlink data channel, which constitutes an NB-IoT downlink frame. NBPU should be embedded avoiding Narrow-band Reference Signal (NRS), subcarriers with fixed phases, in the 5, 6, 12, and 13th symbols of one subframe. By two NRSs in one symbol, only $\binom{10}{2} = 45$ out of a total $\binom{12}{2} = 66$ harmonics can be controlled, reducing the number of controllable by 68\%. NBPU frame with 10 symbols is embedded in the blue box of Figure~\ref{fig:reverse_engineering} while the Rx will ignore the 5 and 6th symbols.

The payload bits that can embed NBPU should be found through reverse engineering of the PHY block diagram in Figure~\ref{fig:reverse_engineering}. In this figure, coded bits $\mathbf{Y}$ and modulated bits $\mathbf{Y}'$ are one-to-one mapping, so its reverse engineering is straightforward. On the other hand, in the 1/3 code rate channel coding, payload bits $\mathbf{X}$ cannot generate arbitrary $\mathbf{Y}$ since the length of $\mathbf{X}$ is shorter than $\mathbf{Y}$ which induces a reduced degree of freedom. To solve this problem, the 240 bits of NBPU can be completely reverse-engineered by adding redundancy bits to $\mathbf{Y}$ such that $\mathbf{X}$ becomes greater than 240 bits. Since the length of payload bits is pre-defined in TS 36.213~\cite{36213}, we utilize a 256-bit payload, the shortest length among candidates, whose result of PHY block diagram is allocated in 3 subframes.

\vspace{-3mm}
\section*{Appendix B. Sweeping NBPU Symbol} \label{sec:appendix_b}
\vspace{-3mm}

For equivalent time sampling, \ours samples with 260ns shifted sampling offset every NB-IoT symbol where symbol length is $66.67\mu$s $\simeq 256\times260$ns. Since the number of symbols transmitted in 20ms is 280 symbols, the accumulated sampling offset after 20ms is $(280-256-1)\times260$ns where $-1$ is for the overflowed sampling offset than symbol length. Thus, after NBPU is transmitted N times (i.e., $N\times20$ms), \ours samples NBPU with $(13\times N \mod 256)\times260$ns sampling offset. Then, the minimum N is 256 to sweep NBPU with 260ns fully. Considering that \ours transmits 5 NBPU symbols every 20ms, we can achieve a fully swept symbol 5 times faster (i.e., N becomes 60). Then, the corresponding time is 1.2s ($=60\times20$ms).

\end{document}